\documentstyle[11pt,epsfig,citesort]{article}

\advance \topmargin by -\headheight
\advance \topmargin by -\headsep
     
\evensidemargin \oddsidemargin
\makeatletter

\def\@sect#1#2#3#4#5#6[#7]#8{\ifnum #2>\c@secnumdepth
     \def\@svsec{}\else 
     \refstepcounter{#1}\edef\@svsec{\csname the#1\endcsname.\hskip 1em }\fi
     \@tempskipa #5\relax
      \ifdim \@tempskipa>\z@ 
        \begingroup #6\relax
          \@hangfrom{\hskip #3\relax\@svsec}{\interlinepenalty \@M #8\par}
        \endgroup
       \csname #1mark\endcsname{#7}\addcontentsline
         {toc}{#1}{\ifnum #2>\c@secnumdepth \else
                      \protect\numberline{\csname the#1\endcsname}\fi
                    #7}\else
        \def\@svsechd{#6\hskip #3\@svsec #8\csname #1mark\endcsname
                      {#7}\addcontentsline
                           {toc}{#1}{\ifnum #2>\c@secnumdepth \else
                             \protect\numberline{\csname the#1\endcsname}\fi
                       #7}}\fi
     \@xsect{#5}}

\renewcommand{\section}{\setcounter{equation}{0} \@startsection {section}{1}
   {\z@}{-3.5ex plus -1ex minus -.2ex}{2.3ex plus .2ex}{\Large\bf}}
\newcommand{\fcaption}[1]{
        \refstepcounter{figure}
        \setbox\@tempboxa = \hbox{\small Fig.~\thefigure. #1}
        \ifdim \wd\@tempboxa > 6in
           {\begin{center}
        \parbox{6in}{\small\baselineskip=12pt Fig.~\thefigure. #1}
            \end{center}}
        \else
             {\begin{center}
             {\small Fig.~\thefigure. #1}
              \end{center}}
        \fi}
\newcommand{\tcaption}[1]{
        \refstepcounter{table}
        \setbox\@tempboxa = \hbox{\small Table~\thetable. #1}
        \ifdim \wd\@tempboxa > 5.4in
           {\begin{center}
        \parbox{5.4in}{\small\baselineskip=12pt Table~\thetable. #1}
            \end{center}}
        \else
             {\begin{center}
             {\small Table~\thetable. #1}
              \end{center}}
        \fi}
 
\title{}
\date{}
\def\a{\alpha}
\def\b{\beta}
\def\d{\delta}
\def\g{\gamma}
\def\l{\lambda}
\def\s{\sigma}

\def\mw{M_W}

\def\beq{\begin{equation}}
\def\eeq{\end{equation}}
\def\bea{\begin{eqnarray}}
\def\eea{\end{eqnarray}}

\newcommand{\im}{\mathop{\mathrm{Im}}}
\newcommand{\re}{\mathop{\mathrm{Re}}}

\begin{document}
\begin{titlepage}

\begin{flushright}
hep-ph/9612380\\
Freiburg--THEP 96/23\\
December 1996
\end{flushright}
\vspace{1.5cm}

\begin{center}

{\LARGE\bf Electroweak ${\cal O}(\a)$  Corrections to $W^+W^-$ Pair\\[.2cm]
       Production in Polarized $\gamma\gamma$ Collisions}

\vspace*{1cm}

{\large\bf\rm G. Jikia\footnote{Alexander von Humboldt Fellow;
	e-mail: jikia@phyv4.physik.uni-freiburg.de}}\\[.5cm]

{\large\bf\em Albert--Ludwigs--Universit\"{a}t Freiburg,
           Fakult\"{a}t f\"{u}r Physik}\\
{\large\bf\em Hermann--Herder Str.3, D-79104 Freiburg, Germany}\\[.5cm]
{\rm and}\\[.5cm]
{\large\bf\em 	Institute for High Energy Physics}\\
{\large\bf\em	Protvino, Moscow Region 142284, Russia}\\[1.5cm]
\end{center}
\normalsize

\begin{abstract}

Full ${\cal O}(\a)$ electroweak corrections for $\g\g\to W^+W^-$ are
calculated including virtual corrections, soft and hard real photon
and $Z$-boson emission. The corrections are quite large ranging from
$-2.5\%$ at 500~GeV to $-18\%$ at 2~TeV,
$|\cos\theta^\pm|<\cos(10^\circ)$. Contributions from real photon
and $Z$-boson emission are important at high energies where they
partly cancel large negative contribution originating from virtual
bosonic corrections.  Precision measurements that intend to uncover
physics beyond the standard model must necessarily make use of the
full standard model predictions including ${\cal O}(\a)$ corrections.

\end{abstract}

\end{titlepage}

\section{Introduction}

A high-energy $\g\g$ linear collider, that can be obtained from
Compton backscattered intense laser beams off electron linac beams
\cite{gg,balakin,schulte} could be a very useful tool for studying the
electro-weak symmetry breaking mechanism which is known as one of the
last most important untested ingredients of the Standard Model (SM).
In models with a light Higgs particle, one unique opportunity for
$\g\g$ colliders relates to the production of the Higgs on
resonance and its ability to perform a direct measurement of the
$H\g\g$ coupling in the reactions $\g\g\to H\to b\bar b,\, ZZ$ \cite{ggH}.
Another very interesting question concerns the observability of the
Higgs boson signal in the reaction $\g\g\to WW$. In the case of very
heavy mass Higgs boson $m_H\ge 1$~TeV, unfortunately, a huge
background from transverse $W_TW_T$ continuum makes searches of the
heavy Higgs signal in $\g\g\to W^+W^-$ hopeless
\cite{higgs1,higgs2,helene}.  However, for $m_H\sim (200\div 300)$~GeV
recently \cite{truong} it was mentioned that on the Higgs resonance
the Born cross section $\g\g\to W^+W^-$, Breit-Wigner Higgs
production cross section and the interference between these two
processes are of the same order of magnitude in $\a$.  Moreover, a
large destructive interference was found between the continuum and the
$s$-channel Higgs exchange, so that Higgs boson would be manifested as
a resonant dip in the $WW$ invariant mass distribution. So, at least
in principle, with excellent $WW$-mass resolution and high luminosity
it will be possible to look for the signal of relatively light Higgs
boson in this reaction.

If no light Higgs boson would be discovered at LEP2, LHC or the linear
collider, the best strategy to probe the symmetry breaking sector
would lie in the study of the self couplings of the $W$, $Z$ bosons.
The large cross sections for the processes involving $W$'s give $\g\g$
colliders an advantage in probing the self couplings.  Indeed, the
reaction $\g\g\to W^+W^-$ would be the dominant source of the $W^+W^-$
pairs at future linear colliders, provided that photon-photon
collider option will be realized. The Born cross section of $W^+W^-$
pair production in photon-photon collisions in the scattering angle
interval $10^\circ < \theta^\pm < 170^\circ$ is 61~pb at
$\sqrt{s_{\g\g}}=500$~GeV and 37~pb at 1~TeV.  Corresponding cross
sections of $W^+W^-$ pair production in $e^+e^-$ collisions are an
order of magnitude smaller: 6.6~pb at 500~GeV and 2.5~pb at
1~TeV. With more than a million $WW$ pairs per year a photon-photon
collider can be really considered as a $W$-factory and an ideal place
to conduct precision tests on the anomalous triple
\cite{yehudai,choi-schrempp,BBB} and quartic \cite{BeBu,BBB,BB}
couplings of the $W$ bosons.  In addition, in the process of triple
$WWZ$ vector boson production it is possible to probe the tri-linear
$ZWW$ and quartic couplings \cite{BB,eboli,BBB,gamma-gamma2} 
as well as the $C$ violating anomalous $ZWW$, $\g ZWW$ interactions
\cite{gamma-gamma2}.

With the natural order of magnitude on anomalous couplings
\cite{Fawzi-Talks}, one needs to know the SM cross sections with a
precision better than 1\% to extract these small numbers. From a
theoretical point of view this calls for the need to calculate full
$\cal{O}(\alpha)$ cross section of $W^+W^-$ pair production in $\g\g$
collisions including radiative corrections and real photon and
$Z$-boson emission.  

In this paper we summarize our results
\cite{gamma-gamma1,gamma-gamma2} for the complete ${\cal O}(\alpha)$
electroweak corrections taking into account one-loop electroweak virtual
radiative corrections, soft and hard photon and $Z$-boson emission.
Virtual and soft-photonic corrections to the process $\g\g\to W^+W^-$
have been also recently calculated by Denner, Dittmaier and Schuster
\cite{denner}. We find complete agreement with the numerical
results of Refs. \cite{denner}.

The paper is organized as follows: In Section~2 we define the helicity
amplitudes, discuss their symmetry properties and define all the
independent helicity amplitudes. In Section~3 the tree-level cross
sections for various polarizations in the high energy limit are
considered. The general structure of ${\cal O}(\alpha)$ corrections is
defined in Section~4. In Section~5 explicit analytical results for the
leading contributions from heavy Higgs boson and top-quark are
given. Also light fermion contributions and asymptotic behavior in
the leading logarithmic approximation at high energy are discussed.
Numerical results are presented in Section~6.

\section{The $\g\g\to W^+W^-$ helicity amplitudes}

The helicity amplitudes for the reaction

\beq
\g(p_1,\l_1)+\g(p_2,\l_2)\to W^+(p_3,\l_3)+W^-(p_4,\l_4)
\eeq
are defined as follows
\beq
{\cal M}_{\l_1\l_2\l_3\l_4}
 = e_1^{\mu_1}(\l_1)e_2^{\mu_2}(\l_2)
{e_3^{\mu_3}}^*(\l_3){e_4^{\mu_4}}^*(\l_4)
G_{\mu_1\mu_2\mu_3\mu_4}(p_1,p_2,p_3,p_4),
\eeq
where $\l_i$ denotes the polarization of particle $i$.
The momenta and polarization vectors for different helicities in the c.m.s. of
the initial photons are given by
\bea
&p_{1,2} = E (1;0,0,\pm 1), & \label{mom}\\ 
&p_{3,4} = -E (1;\pm\b \sin \theta_W, 0, \pm\b \cos \theta_W);&\nonumber
\eea
\bea
&e_1^{\pm} = e_2^{\mp} = \frac{1}{\sqrt{2}}(0;\mp 1,-i,0),&
\nonumber \\
&{e_3^{\pm}}^* ={e_4^{\mp}}^* = 
\frac{1}{\sqrt{2}}(0;\mp \cos \theta_W,+i,\pm \sin \theta_W),&
\label{pol}
\\ &e_{3,4}^0 = \frac{E}{M_W}(\b;\pm\sin\theta_W,0,\pm\cos\theta_W).&
\nonumber \eea All momenta are taken to be incoming.  Here $e^\pm$ is
$\g$ or $W$-boson polarization vector with transverse helicity $\pm 1$
and $e^0$ is the polarization vector of the longitudinal $W$ boson.

The helicity amplitudes are given by the sum of parity-even
($\cal{A}$) and parity-odd ($\cal{B}$) contributions. As bosonic
vertices are $P$ and $C$ even, $\cal{B}$-contribution comes solely
from fermion loop contribution.  \beq {\cal
M}_{\l_1\l_2\l_3\l_4}(s,t,u,\b,p_T) = {\cal
A}_{\l_1\l_2\l_3\l_4}(s,t,u,\b,p_T) + {\cal
B}_{\l_1\l_2\l_3\l_4}(s,t,u,\b,p_T).  \eeq 

We show explicitly $\b$ and $p_T=E\b\sin\theta_W$ as the arguments of
the amplitudes. Taking into account the relations $\b^2=1-4M_W^2/s$
and $p_T^2 = (tu-M_W^4)/s$ the amplitudes can be expressed as rational
functions of Mandelstam variables $s$, $t$ and $u$ \beq
s=(p_1+p_2)^2,\qquad t=(p_2+p_3)^2,\qquad u=(p_1+p_3)^2.  \eeq and
linear function of $\b$, $p_T$ and scalar loop four-, three-, two- and
one-point functions $D$, $C$, $B$ and $A$ \cite{hooft}. With
definitions (\ref{mom})--(\ref{pol}) the following relations take place

\bea 
&& {\cal M}_{\l_1\l_2\l_3\l_4}(s,t,u,\b,-p_T) = (-1)^{\l_3+\l_4}{\cal
M}_{\l_1\l_2\l_3\l_4}(s,t,u,\b,p_T),\label{symm0} \\ && {\cal
M}_{\l_1\l_2\l_3\l_4}(s,t,u,-\b,p_T) = (-1)^{\l_3+\l_4}{\cal
M}_{\l_1\l_2-\l_3-\l_4}(s,t,u,\b,p_T).  \eea

The helicity amplitudes are related by Bose symmetry, parity and charge
conjugation 
\bea
&\mbox{Bose:}&	{\cal M}_{\l_1\l_2\l_3\l_4}(s,t,u,p_T) 
= {\cal M}_{\l_2\l_1\l_3\l_4}(s,u,t,-p_T),
\\
&\mbox{P:}& 	{\cal A}_{\l_1\l_2\l_3\l_4}(s,t,u) 
= (-1)^{\l_3+\l_4}{\cal A}_{-\l_1-\l_2-\l_3-\l_4}(s,t,u),
\\
&\mbox{C:}& 	{\cal A}_{\l_1\l_2\l_3\l_4}(s,t,u,p_T) 
= {\cal A}_{\l_1\l_2\l_4\l_3}(s,u,t,-p_T).\label{symm}
\eea

$P$ and $C$ violating amplitudes $\cal{B}$ acquire an additional minus
sign under parity and charge conjugation transformations.  At
one-loop level $CP$ is an exact symmetry of all the amplitudes even
with complex Cabibbo-Kobayashi-Maskawa matrix.  Taking into account
the symmetry properties (\ref{symm0})--(\ref{symm}) all helicity
amplitudes can be expressed through eight independent $P$-even
helicity amplitudes ${\cal A}$ and six independent $P$-odd helicity
amplitudes ${\cal B}$

\bea
&&{\cal A}_{++++}(\b) = {\cal A}_{++--}(-\b);
\nonumber \\
&&{\cal A}_{+++-} = {\cal A}_{++-+};
\nonumber \\
&&{\cal A}_{+++0}(\b) = {\cal A}_{++0+}(\b) = 
- {\cal A}_{++0-}(-\b) = - {\cal A}_{++-0}(-\b);
\nonumber \\
&&{\cal A}_{++00};
\nonumber \\
&&{\cal A}_{+-++} =  {\cal A}_{+---};
\nonumber \\
&&{\cal A}_{+-+-}(\b) = {\cal A}_{+--+}(-\b);
\nonumber \\
&&{\cal A}_{+-+0}(\b) = {\cal A}_{+-0+}(-\b) =
- {\cal A}_{+-0-}(\b) = - {\cal A}_{+--0}(-\b);
\nonumber \\
&&{\cal A}_{+-00};
 \\
&&{\cal B}_{++++} = {\cal B}_{++--} = {\cal B}_{++00} = 0;
\nonumber \\
&&{\cal B}_{+++-}(\b) = {\cal B}_{++-+}(-\b);
\nonumber \\
&&{\cal B}_{+++0}(\b) = -{\cal B}_{++0+}(\b) =
{\cal B}_{++0-}(-\b) = - {\cal B}_{++-0}(-\b);
\nonumber \\
&&{\cal B}_{+-++} = {\cal B}_{+---};
\nonumber \\
&&{\cal B}_{+-+-}(\b) = {\cal B}_{+--+}(-\b);
\nonumber \\
&&{\cal B}_{+-+0}(\b) = {\cal B}_{+-0+}(-\b) = 
- {\cal B}_{+-0-}(\b)  = - {\cal B}_{+--0}(-\b);
\nonumber \\
&&{\cal B}_{+-00}.
\nonumber 
\eea

It is worth mentioning that tree level amplitudes are equal to zero for equal
helicities of incoming photons and unequal helicities of final $W$'s
\beq
{\cal M}^{Born}_{++LT}={\cal M}^{Born}_{++TL}=
{\cal M}^{Born}_{+++-}={\cal M}^{Born}_{++-+}=0.
\eeq
It follows that corresponding one-loop amplitudes, which are not zero,
do not contribute to the interference term if the cross section is
integrated over the phase space of the $W$'s decay products. However,
they do contribute if angular distributions and correlations of the
decay products are reconstructed.

\section{Born Cross Sections}

The lowest order polarized cross sections at high energy $s\gg M_W^2$
are given by (see also the full Born helicity amplitudes
and cross sections \cite{yehudai,BeBu,truong})

\bea
&& \int\limits_{{ p_T^W>p_T}}\!\!\!\!\!\!
{ d\s^{Born}_{++++}}={\frac {8\,\pi \,{\alpha}^{2}}{ p_T^2}}
\,+\,\cdots \: \mathop{\longrightarrow}_{{ p_T}\to 0} \: 
                    {\frac {8\,\pi \,{\alpha}^{2}}{M_W^{2}}}
\nonumber \\
&& \int\limits_{{ p_T^W>p_T}}\!\!\!\!\!\!
{ d\s^{Born}_{++00}}={\frac {8\,\pi \,{\alpha}^{2}M_W^{4}}
{{p_T^2}{s}^{2}}}
\,+\,\cdots \: \mathop{\longrightarrow}_{{ p_T}\to 0} \: 
                    {\frac {8\,\pi \,{\alpha}^{2}M_W^{2}}{{s}^{2}}}
\nonumber \\
&& \int\limits_{{ p_T^W>p_T}}\!\!\!\!\!\!
{ d\s^{Born}_{++--}}={\frac {8\,\pi \,{\alpha}^{2}M_W^{8}}
{{p_T^2}{s}^{4}}}
\,+\,\cdots \: \mathop{\longrightarrow}_{{ p_T}\to 0} \: 
                    {\frac {8\,\pi \,{\alpha}^{2}M_W^{6}}{{s}^{4}}}
\nonumber \\
&& \int\limits_{{ p_T^W>p_T}}\!\!\!\!\!\!
{ d\s^{Born}_{+-+-}}={\frac {4\,\pi \,{\alpha}^{2}}{ p_T^2}}
\,+\,\cdots \: \mathop{\longrightarrow}_{{ p_T}\to 0} \: 
                    {\frac {4\,\pi \,{\alpha}^{2}}{M_W^{2}}}
\\
&& \int\limits_{{ p_T^W>p_T}}\!\!\!\!\!\!
{ d\s^{Born}_{+-00}}={\frac {4\,\pi \,{\alpha}^{2}}{s}}\,+\,\cdots
\nonumber \\
&& \int\limits_{{ p_T^W>p_T}}\!\!\!\!\!\!
{ d\s^{Born}_{+-+0}}=\frac {32\,\pi \,{\alpha}^{2}M_W^{2}}{s^2}
\left (\ln \frac{s}{{ p_T^2}}\,-\,1\right )
\,+\,\cdots \: \mathop{\longrightarrow}_{{ p_T}\to 0} \: 
                    \frac {32\,\pi \,{\alpha}^{2}M_W^{2}}{s^2}
\left (\ln {\frac {s}{M_W^{2}}}\,-\, 2 \right )
\nonumber \\
&& \int\limits_{{ p_T^W>p_T}}\!\!\!\!\!\!
{ d\s^{Born}_{+-++}}={\frac {64\,\pi \,{\alpha}^{2}M_W^{4}}{{s}^
{3}}}\,+\,\cdots
\nonumber
\eea

The first column shows the cross sections integrated over a region
$p_T^W>p_T$, for $s\gg p_T^2\gg M_W^2$. The last column shows the
cross sections integrated over the whole phase space. For initial
helicities $+-00$ and $+-++$ these two limit coincide. As is well
known (see {\it e.g.} \cite{Lipatov}), the asymptotic behavior of the
scattering amplitudes is determined by the exchange of the spin-one
particle in the $t$-channel and in the Born approximation is given by
\beq 
{\cal M}_{\l_1\l_2\l_3\l_4}\Biggl|_{\begin{array}{c}
s\to \infty\\
t\sim M_W^2
\end{array}} \propto \frac{2\,s}{t-M_W^2}
\d_{\l_1\l_4}\d_{\l_2\l_3},
\eeq
so that the cross sections integrated over the whole phase space are
non-decreasing with energy for helicity conserving
amplitudes. Asymptotically 
\beq
\s^{Born}_{+-+-}=\s^{Born}_{+--+}=\frac{1}{2}\s^{Born}_{++++}, 
\eeq
and total cross sections of $WW$-pair production are the same for $++$
and $+-$ initial photon helicities.

As initial photons are necessarily transversely polarized, cross
section of transverse $W_TW_T$ pair production dominates. For a finite
angular cut even the dominating cross sections do decrease as $1/s$,
but still they are much larger than the suppressed cross sections
because they are still enhanced by a large factor of $s/p_T^2$. A
cross section $\s_{+-00}$ is the next to the largest ones $\s_{++++}$,
$\s_{+-+-}$ and it is decreasing as $1/s$. The other cross sections
decrease for finite angular cuts as $1/s^2$ ($\s_{+-+0}$), $1/s^3$
($\s_{++00}$, $\s_{+-++}$), and even as $1/s^5$ ($\s_{++--}$). The
cross sections $\s_{++LT}$ and $\s_{+++-}$ vanish at the Born level.

\section{${\cal O}(\a)$ corrections}

Inclusive cross section of $W^+W^-$ pair production in photon-photon
collisions to third order in $\a$ is given by the sum of Born cross
section, interference term between the Born and one-loop amplitudes
and cross section of $W^+W^-$ pair production accompanied by the real
photon emission. At energies above the $WWZ$ threshold one should add
the cross section of $W^+W^-Z$ production. In principle one can also
add the cross section of associated Higgs boson production $\g\g\to
W^+W^-H$. It seems however unreasonable to consider here associated
Higgs production on an equal footing with gauge boson production, as
it definitely deserves a special consideration \cite{BB,WWWW}. Anyway,
since the cross section of $W^+W^-H$ production is at least an order
of magnitude smaller than that of $W^+W^-Z$, its contribution to
corrections to $\g\g\to W^+W^-$ cross section can be safely neglected.
Since one-loop amplitude and soft photon emission cross section are
IR-divergent and only their sum is IR-finite it is convenient, as
usual, to consider soft and hard photon emission separately. Finally,
the inclusive cross section of $W^+W^-$ pair production in
photon-photon collisions can be represented as a sum 
\bea
&&d\s(\g\g\to W^+W^-+X)\,=\, 
\label{cs}\\
&&d\s^{Born}(\g\g\to W^+W^-)\,+\,
\frac{1}{2s_{\g\g}} 2 \re\Biggl({\cal M}^{Born}
{{\cal M}^{1-loop}}^*\Biggr) dPS^{(2)}
\nonumber \\
&&+\,d\s^{soft}(\g\g\to W^+W^-\g)\biggr|_{\omega_\g<k_c}
\,+\,d\s^{hard}(\g\g\to W^+W^-\g)\biggr|_{\omega_\g>k_c}
\,+\,d\s^Z(\g\g\to W^+W^-Z).
\nonumber
\eea

Analytical expression in terms of Spence functions for the
IR-divergent factorizable soft photon emission cross section is given
by 
\bea 
d\s^{soft}(\g\g\to W^+W^-\g) = d\s^{Born}(\g\g\to W^+W^-)\,R^{soft}, 
\eea 
where 
\bea R^{soft} &=& \frac{2\alpha}{\pi}\Biggl\{
\left(-1+\frac{1}{\b}(1-\frac{2 \mw^2}{s})
\ln\biggl(\frac{1+\b}{1-\b}\biggr)\right)
\ln\biggl(\frac{2k_c}{\l}\biggr) \\
&&+\frac{1}{2\b}\ln\biggl(\frac{1+\b}{1-\b}\biggr)+
\frac{1}{2\b}\biggl(1-\frac{2 \mw^2}{s}\biggr)
\Biggl[Sp\biggl(\frac{-2\b}{1-\b}\biggr)
-Sp\biggl(\frac{2\b}{1+\b}\biggr)\Biggr]\Biggr\} \nonumber 
\eea
and $\l$ is fictitious photon mass, $k_c$ is the soft photon energy
cutoff and $\b=\sqrt{1-4\mw^2/s}$.

The cross section of hard photon emission is given by
\beq
\s^{hard}_{\l_1\l_2\l_3\l_4}(\g\g\to W^+W^-\g) 
= \frac{1}{2s_{\g\g}}\int \sum_{\l_5}
|{\cal M}_{\l_1\l_2\l_3\l_4\l_5}(\g\g\to W^+W^-\g)|^2\, dPS^{(3)},
\eeq
where the three particle phase space can be represented in the following form
\beq
dPS^{(3)} = \frac{1}{4(2\pi)^3}\omega_\g^2\b_{+-}\,
 d\ln\omega_\g\, d\cos\chi\, d\cos\theta^+\, d\cos\theta^-.
\eeq
Here $\omega_\g$ is the final photon energy in c.m.s. of colliding
initial photons, $\chi$ is the final photon -- $W$ angle in the
c.m.s. of $W^+W^-$ system, $\theta^\pm$ are the initial photon --
$W^\pm$ angles in c.m.s. of colliding photons and
$\b_{+-}=\sqrt{1-4M_W^2/s_{W^+W^-}}$. The cross section of $W^+W^-Z$
production is given by the analogous formulas.

Total ${\cal O}(\a)$ correction is defined by
\beq
\s(\g\g\to W^+W^-+X) = \s^{Born}(\g\g\to W^+W^-)\,(1+\d^{tot}).
\eeq
According to (\ref{cs}) it can be decomposed as
\beq
\d^{tot}=\d^{virt}\,+\,\d^{soft}\,+\,\d^{hard}\,+\,\d^{Z}.
\eeq

On-shell renormalization scheme \cite{OS} and 't~Hooft-Feynman
gauge are used to calculate loop corrections, {\it i.e.}
fine-structure constant $\alpha$ and the physical particle masses are
used as basic parameters.  The width of the Higgs boson resonance is
included in a gauge invariant manner, as described {\it e.g.} in
Ref. \cite{stuart}.

The calculation has been performed using symbolic manipulation program
{\it FORM} \cite{FORM} to generate Feynman diagrams and to reduce
tensor loop integrals to scalar one-loop one-, two-, three- and
four-point functions $A$, $B$, $C$ and $D$
\cite{hooft,reduction}. $FF$-package \cite{FF} has been used for the
numeric evaluation of the scalar loop integrals. The $\g\g\to
W^+W^-\g$, $W^+W^-Z$ helicity amplitudes have also been evaluated with
{\it FORM}. For the phase space integration we have used the
multi-dimensional Monte Carlo integration package $BASES$
\cite{BASES}.

\section{The structure of leading corrections}

As it was already mentioned in the literature
\cite{helene,higgs2,denner}, no power or logarithmically enhanced
corrections involving $m_H^2/M_W^2$, $\log(m_H^2)$, $m_t^2/M_W^2$ or
$\log(m_t^2)$ appear in the on-shell scheme in the limit, when
Higgs-boson or top-quark masses are much larger than the collision
energy.  The explanation of this fact is given by the low energy
theorem, according to which the cross section of the Compton
scattering of the photon off the $W$-boson target, when photon energy
tends to zero, to all orders is given by the Born cross section with
the universally renormalized on-shell electric charge \cite{OS}
\beq 
e^R = \Biggl(Z^{1/2}_{AA} - \frac{s^0_W}{c^0_W}Z^{1/2}_{ZA}\Biggr)\,e^0.
\label{universal}
\eeq 
Here the charge universality means that electric charge measured in the 
photon scattering off $W$-boson is equal to the charge measured in 
photon-electron Compton scattering. Since large masses $m_H$, $m_t$ can be
considered as gauge invariant cutoffs, divergent in the limit 
$m_H^2,m_t^2\gg s,-t,-u$ contributions should take a form of local gauge 
invariant three- and four-vertex corrections, {\it i.e.} their contribution
to the amplitude of the Compton photon-$W$ scattering should be proportional
to the Born amplitude. Moreover, this contribution should be canceled after
the counterterms are taken into account in order to guarantee the charge
universality.

Indeed, if we include only the tadpole and $W$-boson 
mass counterterms, to fix a mass of the $W$-boson, the divergent in the heavy
mass limit amplitudes are given by
\bea
&& {\cal M}^{bose}_{div} = -\frac{\alpha}{48\pi s_W^2}\log(m_H^2)
{\cal M}^{Born},\\
&& {\cal M}^{fermi}_{div} = -\frac{\alpha N_c}{12\pi s_W^2}
\log(m_t^2) {\cal M}^{Born}.
\eea
As soon as the
$W$-boson wave function renormalization constant is taken into account these
divergent contributions are really canceled out (renormalization constants
$Z^{1/2}_{AA}$, $Z^{1/2}_{ZA}$ contain no logs or positive powers of
the ratios $m_H^2/M_W^2$, $m_t^2/M_W^2$ at one-loop level).

Instead of rising with $m_H$, $m_t$ corrections, rising with
energy contributions still remain \cite{higgs1,higgs2} as a result of
violation of unitarity cancellations in this limit. Namely, in the
limit $M_W^2\ll s,-t,-u \ll m_H^2$ the following helicity amplitudes
exhibit rising contributions

\bea
&& {\cal M}^{bose}_{++00} = \frac{5\alpha^2}{12 s_W^2}\frac{s}{M_W^2},
\\
&& {\cal M}^{bose}_{+++0} = 
\frac{\alpha^2}{6 \sqrt{2} s_W^2}\frac{t-u}{M_W p_T},
\\
&& {\cal M}^{bose}_{+-+0} = -\frac{\alpha^2}{6 \sqrt{2} s_W^2}
\frac{t}{M_W p_T}.
\eea
The heavy top-quark contributions in the limit
$M_W^2\ll s,-t,-u \ll m_t^2$ are given by
\bea
&& {\cal M}^{fermi}_{++++} = {\cal M}^{fermi}_{++--} = 
-\frac{4\alpha^2 N_c}{27 s_W^2}\frac{s}{s-m_H^2},
\\
&& {\cal M}^{fermi}_{++00} = -\frac{\alpha^2 N_c}{54 s_W^2}\frac{s}{M_W^2}
\,-\, \frac{4\alpha^2 N_c}{27 s_W^2}\frac{s}{s-m_H^2}
\left(\frac{m_H^2}{2M_W^2}-1\right),
\\
&& {\cal M}^{fermi}_{+-+0} = 	\frac{\alpha^2 N_c}{6 \sqrt{2} s_W^2}
				\frac{s-t}{s}
				\frac{t}{M_W p_T }.
\eea

In addition, at energies much larger than $m_H$, $m_t$ power corrections 
proportional to $m_H^2/M_W^2$, $m_t^2/M_W^2$ do appear as a consequence of
incomplete unitarity cancellations. For
$M_W^2\ll m_H^2\ll s,-t,-u $ we have\footnote{The right hand side of equation
(11) of Ref. \cite{higgs2} should be multiplied by 2.}
\bea
&& {\cal M}^{bose}_{++00} = -\frac{\alpha^2}{s_W^2}\frac{m_H^2}{M_W^2},
\label{mh2pp}\\
&& {\cal M}^{bose}_{+-00} = \frac{\alpha^2}{4 s_W^2}\frac{m_H^2}{M_W^2}.
\label{mh2pm}
\eea

The leading top-quark contributions in the limit $M_W^2\ll m_t^2\ll s,-t,-u$
are given by
\bea
&& {\cal M}^{fermi}_{++++} = {\cal M}^{fermi}_{++--}=
 \frac{16 \alpha^2 N_c}{9 s_W^2}\frac{m_t^2}{s-m_H^2},
\label{mt2pppp}\\
&& {\cal M}^{fermi}_{++00} = \frac{5 \alpha^2 N_c}{9 s_W^2}\frac{m_t^2}{M_W^2}
\,+\, \frac{16 \alpha^2 N_c}{9 s_W^2}\frac{m_t^2}{s-m_H^2}
\left(\frac{m_H^2}{2M_W^2}-1\right),
\label{mt2pp00}\\
&& {\cal M}^{fermi}_{+-00} = \frac{\alpha^2 N_c}{s_W^2}\frac{m_t^2}{M_W^2}
\left\{\frac{5}{18}\ \frac{t}{u}\left(\log^2 (\frac{s}{t}-i\epsilon)
+\pi^2\right)+\log(-\frac{t}{m_t^2})-\frac{3}{2}\right\}.
\label{mt2pm00}
\eea

Cross sections of the longitudinal $W_LW_L$ pair production are the
most sensitive to the mechanism of the electroweak symmetry breaking.
Although the dependence is quite strong $\propto m_H^2,\,m_t^2$
(\ref{mh2pp})--(\ref{mt2pm00}), the cross sections $\s_{+\pm 00}$
themselves are quite small in comparison to cross section of
transverse $W_TW_T$ pair production and it will be very difficult to
measure Higgs or top quark couplings with longitudinal week bosons in
this reaction.

Another very important feature of radiative corrections to $\g\g\to
W^+W^-$, also mentioned in the Refs. \cite{denner}, is that no large
logarithms containing light fermion masses $\log(s/m_f^2)$ remain in
the renormalized amplitude, so that not running $\a\approx 1/129$ at
$q^2=M_W^2$, but fine-structure constant $\a=1/137.036$ at zero
momentum is relevant for this reaction. For $\g W$ Compton scattering
near the threshold this property is an immediate consequence of the
low energy theorem and charge universality mentioned above and it
should be true to any order of $\a$.  In fact, at one-loop level one
can easily show, that these large logarithms cancel at any energy.
More specifically, mass singularities coming from the light fermion
contributions to the scalar loop integrals are canceled out in the
total unrenormalized amplitude as a consequence of the
Kinoshita-Lee-Nauenberg (KLN) theorem \cite{KLN}. Due to the identity
(\ref{universal}), singular terms that are present in the photon wave
function renormalization constant $Z^{1/2}_{AA}$ are canceled by the
corresponding terms in the renormalization constant of the electric
charge.  The fermion contribution to constant $Z^{1/2}_{ZA}$ is equal
to zero at one-loop level.

And last but not least, the asymptotic behavior of the dominating
gauge vector boson scattering amplitude in the leading logarithmic 
approximation at high energy and fixed momentum transfers $s\gg -t,\,M_W^2$, 
related to the exchange of spin-one and isospin  $I=1$
particle in the $t$-channel, is  known (see, {\it e.g.} \cite{Lipatov,Wu}) 
to have the Regge form
\beq
{\cal M}_{\l_1\l_2\l_3\l_4}(s,t,u) \Biggr|_{I=1} \propto \frac{g^2}{t-M_W^2}
\d_{\l_1\l_4}\d_{\l_2\l_3}\left((-s)^{1+\a(t)}-(-u)^{1+\a(t)}\right).
\label{Regge}
\eeq
Here the trajectory $\a(t)$ is given by
\bea
&&\a(q^2) = \frac{g^2}{(2\pi)^3}\,(q^2-M_W^2)\,\int\frac{d^2k_\perp}
{\left(k_\perp^2-M_W^2\right)\left((k_\perp+q)^2-M_W^2\right)},
\label{alpha}\\
&&\a(0) = -\frac{g^2}{8\pi^2},\quad
 \a(t)\biggr|_{-t\gg M_W^2} \sim -\frac{g^2}{4\pi^2}\log(-t/M_W^2).
\label{alpha0}
\eea
In fact, the equation (\ref{Regge}) was explicitly checked up to the eights 
order of perturbation theory for the gauge group $SU(2)$ \cite{Lipatov,Wu}.

With the help of the equations (\ref{Regge})--(\ref{alpha0}) one can
easily derive, that total ${\cal O}(\a)$ bosonic one-loop correction to the 
total cross section of vector boson scattering, including virtual corrections 
as well as real triple vector boson production grows as $\log(s/M_W^2)$ and 
is given by
\beq
\d^{tot} 
= \frac{\im {\cal M}^{2-loop}}{\im {\cal M}^{1-loop}}\Biggr|_{t=0}
=\a(0)\log(s/M_W^2) = - \frac{\a}{2 \pi s_W^2} \log(s/M_W^2).
\eeq
So, one can guess that electroweak corrections to the total $\g\g\to W^+W^-$
cross section are negative (because the trajectory $\a(t)$ (\ref{alpha}) is 
negative) and range from $\sim -1\%$ at $\sqrt{s}=300$~GeV to $\sim -3\%$ at
$\sqrt{s}=1$~TeV.

From the other side, one can estimate the value of radiative corrections in
the central region of $W^+W^-$ production by just integrating the one-loop
scattering cross section (\ref{Regge}) in this region
\beq
\d^{virt}(|t|>|t_0|) \approx -2\frac{\a}{\pi s_W^2} 
\log(s/M_W^2) \log(-t_0/M_W^2).
\eeq
So, one should expect that corrections to $\g\g\to W^+W^-$ cross
section in the central region grow as $\log^2(s/M_W^2)$ and could be
$5\div 10$ times larger than corrections to the total cross section at
TeV energies.

\section{Numerical results}

The following set of parameters was used \cite{PDG}

\begin{tabular}{lll}
$\a$ = 1/137.0359895 \\
$M_Z$ = 91.187~GeV, 		& $M_W$ = 80.36~GeV, 	& $m_H$ = 300~GeV \\
$m_e$ = 0.51099907~MeV, 	& $m_u$ = 48~MeV, 	& $m_d$ = $m_u$, \\
$m_\mu$ = 0.105658389~GeV,	& $m_c$ = 1.5~GeV, 	& $m_s$ = 150~MeV, \\ 
$m_\tau$ = 1.777~GeV, 		& $m_t$ = 180~GeV, 	& $m_b$ = 4.5~GeV.
\end{tabular}

\noindent The unit Cabibbo-Kobayashi-Maskawa matrix was used and the
$W$-boson width was neglected.
  
An important issue relates to the choice of the photon spectrum at
$\g\g$ collider. In principle, it is possible to adjust the parameters
of the laser to obtain a spectrum most suitable for gauge boson
production, one that is polarized and peaked at high energies
\cite{gg}.  However, the precise high energy photon spectrum depends
on the beam polarization, conversion distance and other technical
parameters \cite{gg,balakin,schulte} and can vary from one scheme to
the other and the shape of the predicted ideal spectrum \cite{gg} can
be severely distorted \cite{balakin,schulte}. Since the detailed
studies and choices of the optimal parameters have not been made yet,
we prefer to present our results for the case of monochromatic photon
spectrum. As a worst case scenario, the effect of non-monochromaticity
can be estimated with a luminosity in the peak that is ten times
smaller than the total photon-photon luminosity.

Since photon energy distribution in the reaction of $W^+W^-\g$
production strongly peaks at low energies it will not be easy to
experimentally separate $W^+W^-$ pair production from $W^+W^-\g$
production. Moreover, in realistic case the energy of the initial
photons will not be fixed because of the quite wide spectrum of the
Compton backscattered photons \cite{gg,balakin,schulte}, and it will
not be possible to require that $W$-bosons would be produced
back-to-back to suppress real photon emission.  So, in what follows we
just integrate over the whole photon phase space available. Angular
cuts will be imposed only on $W^\pm$ production angles.

\begin{figure}
\setlength{\unitlength}{1in}
\begin{picture}(6,3.5)
\put(-.15,0){\epsfig{file=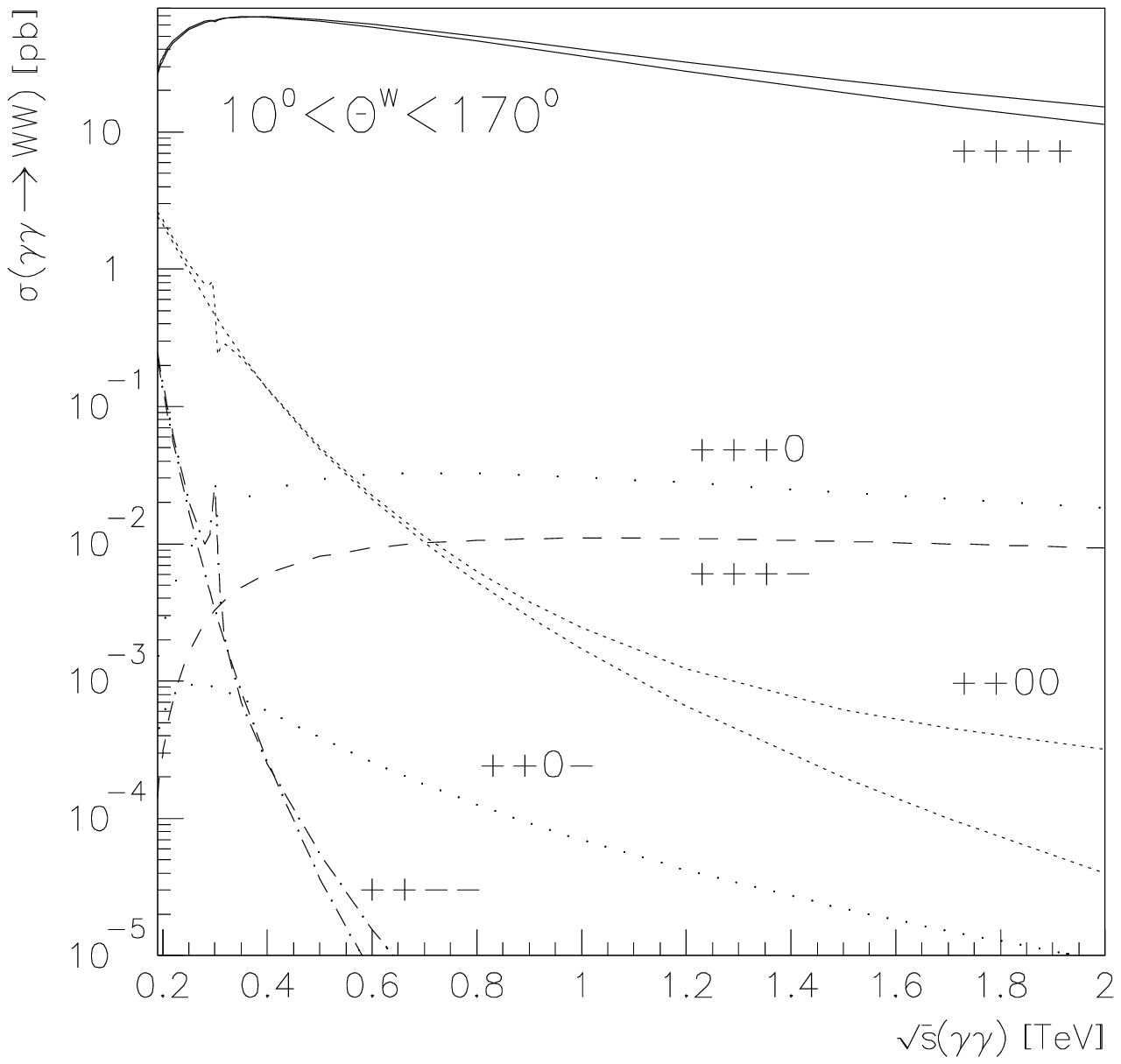,width=3.2in,height=3.5in}}
\put(2.85,0){\epsfig{file=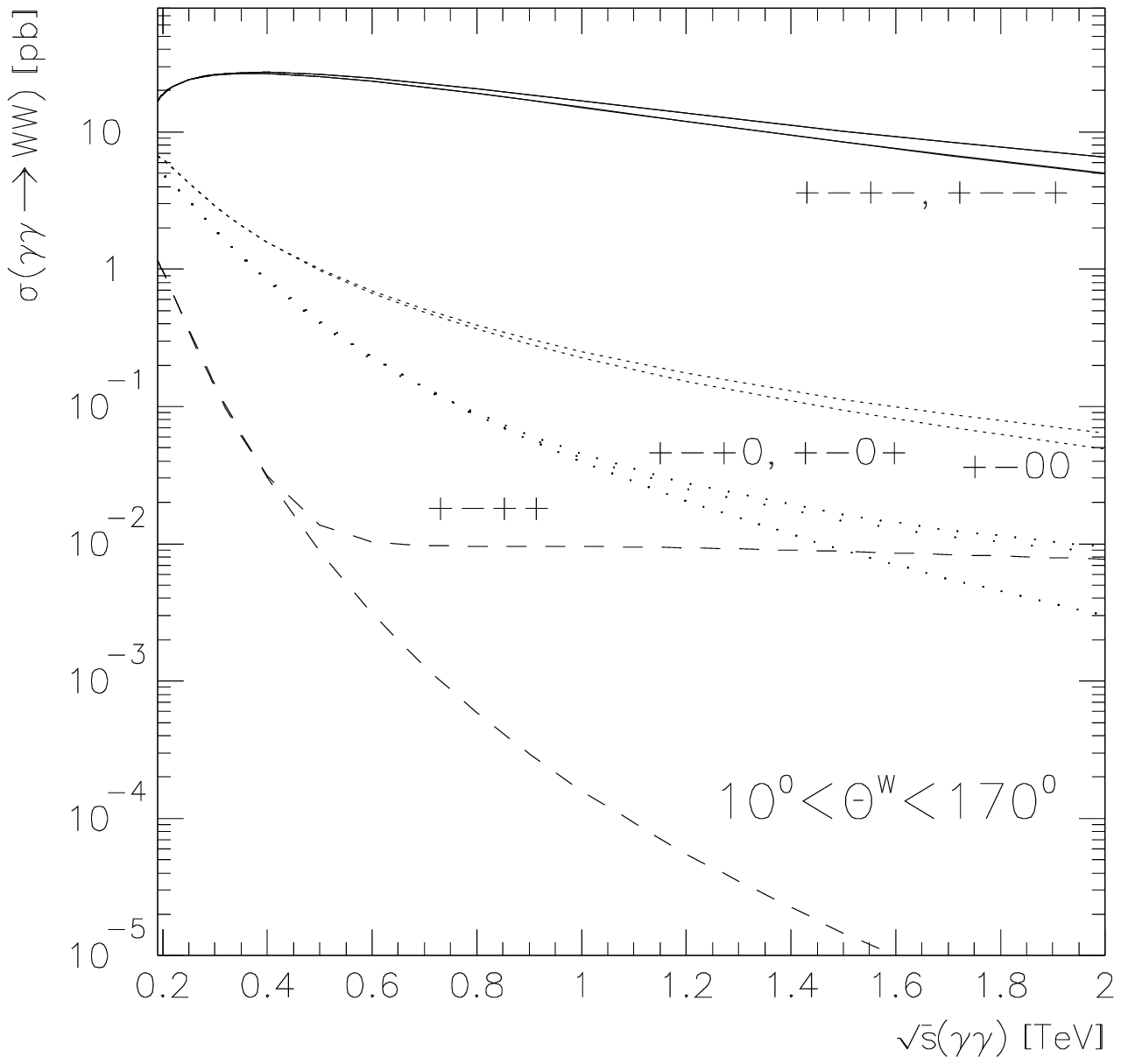,width=3.2in,height=3.5in}}
\end{picture}
\fcaption{Total cross sections of $WW(\g)$ production for various
polarizations.  Born and corrected cross sections are shown. The
curves nearest to the helicity notations represent the corrected cross
sections.}
\end{figure}

\begin{table}
\tcaption{Total Born cross sections and relative corrections for
various polarizations. When Born cross section is equal to zero cross
sections of hard photon emission are given instead of relative
corrections. The $W^\pm$ production angle is restricted to
$10^\circ<\theta^\pm<170^\circ$. $WWZ$ production is not included.}

\begin{center}
$\sqrt{s} = 500$~GeV

\begin{tabular}{|c|c|c|c|c|c|c|}\hline
$\lambda_1\lambda_2\lambda_3\lambda_4$
& $\sigma^{Born}$, $pb$ & $\delta^{hard}$, \% &
$\delta^{soft}$, \% & $\delta^{bose}$, \% & $\delta^{fermi}$, \% 
&$\delta^{tot}$, \% \\ \hline
unpol  &  60.71  &  7.89  & 2.30 & $-$13.0 & $-$0.242 & $-$3.06    
\\ \hline
$++TT$ &  66.12  &  7.91  & 2.30 & $-$13.0 & $-$0.137 & $-$2.93    
\\ \hline
$++TL$ &    0    & 6.01$\cdot 10^{-2}pb$ & 0 & 0 & 0 & --      
\\ \hline
$++LL$ & 4.913$\cdot 10^{-2}$ & 13.7 & 2.30 & $-$22.4 & 9.89 & 3.78    
\\ \hline
$+-TT$ &  52.58  &  7.66  & 2.30 & $-$13.0 & $-$0.309 & $-$3.37    
\\ \hline
$+-TL$ &  1.669  &  10.2  & 2.30 & $-$12.6 & $-$3.06  & $-$3.18    
\\ \hline
$+-LL$ & 0.9989  &  8.20  & 2.30 & $-$12.6 & 0.575  & $-$1.53    
\\ \hline
\end{tabular}
\vspace{.5cm}

$\sqrt{s} = 1000$~GeV

\begin{tabular}{|c|c|c|c|c|c|c|}\hline
$\lambda_1\lambda_2\lambda_3\lambda_4$
& $\sigma^{Born}$, $pb$ & $\delta^{hard}$, \% &
$\delta^{soft}$, \% & $\delta^{bose}$, \% & $\delta^{fermi}$, \% 
&$\delta^{tot}$, \% \\ \hline
unpol  &  37.04  &  13.4 & 3.11 & $-$25.7 & $-$1.28 & $-$10.5    
\\ \hline
$++TT$ &  40.08  &  13.4 & 3.11 & $-$26.4 & $-$1.30 & $-$11.1    
\\ \hline
$++TL$ &    0    & 6.13$\cdot 10^{-2}pb$ & 0 & 0 & 0 & --     
\\ \hline
$++LL$ & 1.715$\cdot 10^{-3}$ &   64.3 &  3.11 & $-$34.8 & 10.3 & 42.9    
\\ \hline
$+-TT$ &  33.60  &  13.1 & 3.11 & $-$25.0 & $-$1.19 & $-$10.0    
\\ \hline
$+-TL$ & 0.1577  &  40.5 & 3.11 & $-$21.9 & $-$10.6 &  11.2    
\\ \hline
$+-LL$ & 0.2510  &  14.2 & 3.11 & $-$24.4 & $-$2.82 & $-$9.86    
\\ \hline
\end{tabular}
\vspace{.5cm}

$\sqrt{s} = 2000$~GeV

\begin{tabular}{|c|c|c|c|c|c|c|}\hline
$\lambda_1\lambda_2\lambda_3\lambda_4$
& $\sigma^{Born}$, $pb$ & $\delta^{hard}$, \% &
$\delta^{soft}$, \% & $\delta^{bose}$, \% & $\delta^{fermi}$, \% 
&$\delta^{tot}$, \% \\ \hline
unpol  &  14.14  &  20.1 & 3.49 & $-$45.1 & $-$2.99 & $-$24.5    
\\ \hline
$++TT$ &  15.17  &  20.0 & 3.49 & $-$46.1 & $-$3.03 & $-$25.6    
\\ \hline
$++TL$ &  0  & 3.65$\cdot 10^{-2}pb$ & 0 &  0 &  0 &  --     
\\ \hline
$++LL$ & 3.979$\cdot 10^{-5}$ & 749. &  3.49 & $-$84.8 & 35.0 & 702.    
\\ \hline
$+-TT$ &  13.04  & 19.6 & 3.49 & $-$44.0 & $-$2.89 & $-$23.8   
\\ \hline
$+-TL$ & 1.197$\cdot 10^{-2}$ & 248. &  3.49 & $-$29.8 & $-$22.2 & 199.    
\\ \hline
$+-LL$ & 6.364$\cdot 10^{-2}$ & 21.0 &  3.49 & $-$40.2 & $-$7.75 & $-$23.4    
\\ \hline
\end{tabular}
\end{center}
\end{table}

Figure~1 shows total cross section of $WW$ pair production summed over
$WW$ and $WW\g$ final states and integrated over $W^\pm$ scattering
angles in the interval $10^\circ<\theta^\pm<170^\circ$ as a function of
energy for various polarizations.  As we have discussed in Sections~3,
5 the bulk of the cross section originates from transverse $W_TW_T$
pair production.  Transverse $W$'s are produced predominantly in the
forward/backward direction and the helicity conserving amplitudes are
dominating. Cross sections integrated over the whole phase space are
non-decreasing with energy. For a finite angular cutoff they do decrease
as $1/s$, but still they are much larger than suppressed cross
sections. For the dominating $++++$, $+-+-$, $+--+$ helicity
configurations corrections are negative and they rise with energy
ranging from $-3\%$ at 500~GeV to $-25\%$ at 2~TeV. The correction for
the next to the largest cross section $\s_{+-00}$ is also
negative. For the other helicities radiative corrections are positive
at high energies due to dominating positive contribution of real
photon emission.  Radiative corrections to cross sections $\s_{+-TL}$,
which are decreasing at Born level as $1/s^2$, are positive and
large. For the cross sections $\s_{++00}$ and $\s_{+-++}$, which are
decreasing as $1/s^3$ for a finite angular cutoff, corrections are even
larger. The cross section $\s_{++--}$ is decreasing at tree level as
$1/s^5$ for a finite angular cutoff and is quite negligible at high
energy.  The cross sections $\s_{++LT}$ and $\s_{+++-}$ vanish at the
Born level, so only the process of $WW\g$ production contributes in
Figure~1.  The cross sections $\s_{++--}$ and $\s_{++00}$ exhibit a clear
clear Higgs resonance peak and the interference pattern,
respectively. The dominating $\s_{++++}$ cross section has a hardly
visible in logarithmic scale dip at $\sqrt{s_{\g\g}}=m_H=300$~GeV
\cite{truong}. A very high $WW$ mass resolution and high luminosity
will be required to register such a dip experimentally.

Table~1 presents Born cross sections and relative corrections for
various polarizations and energies. Total corrections,  corrections
originating from virtual bosonic and fermionic contributions
$\d^{virt}=\d^{bose}+\d^{fermi}$ as well as from soft and hard photon
emission are shown separately. Fictitious photon mass $\l=10^{-2}$~GeV
is used to regularize infrared divergences and a photon energy cutoff
$k_c=0.1$~GeV discriminates between soft and hard
bremsstrahlung. Looking at the Table~1 one concludes that bosonic
corrections are obviously dominating over fermionic ones. 

\begin{figure}
\setlength{\unitlength}{1in}
\begin{picture}(6,7)
\put(-.15,3.5){\epsfig{file=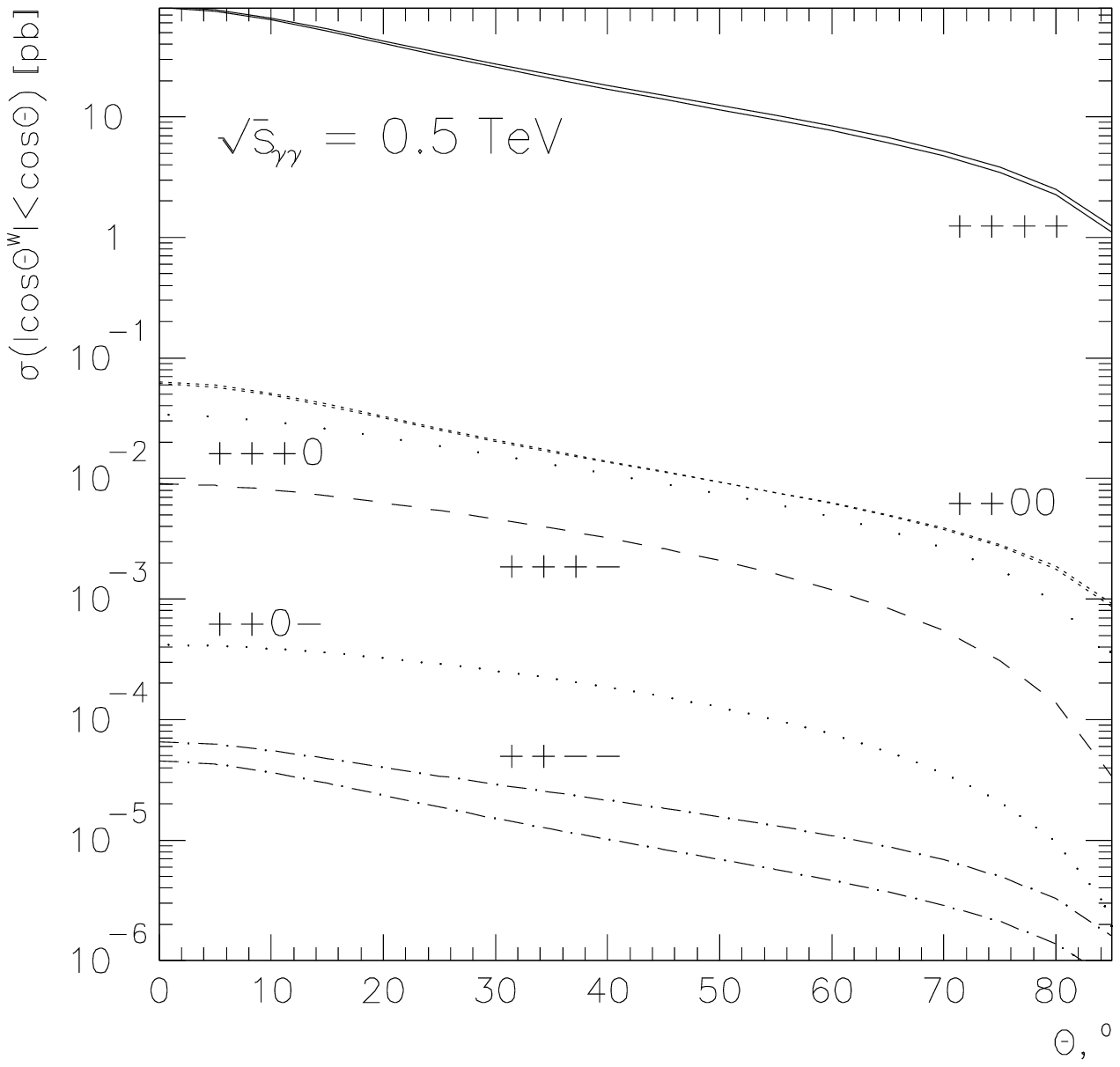,width=3.2in,height=3.5in}}
\put(2.85,3.5){\epsfig{file=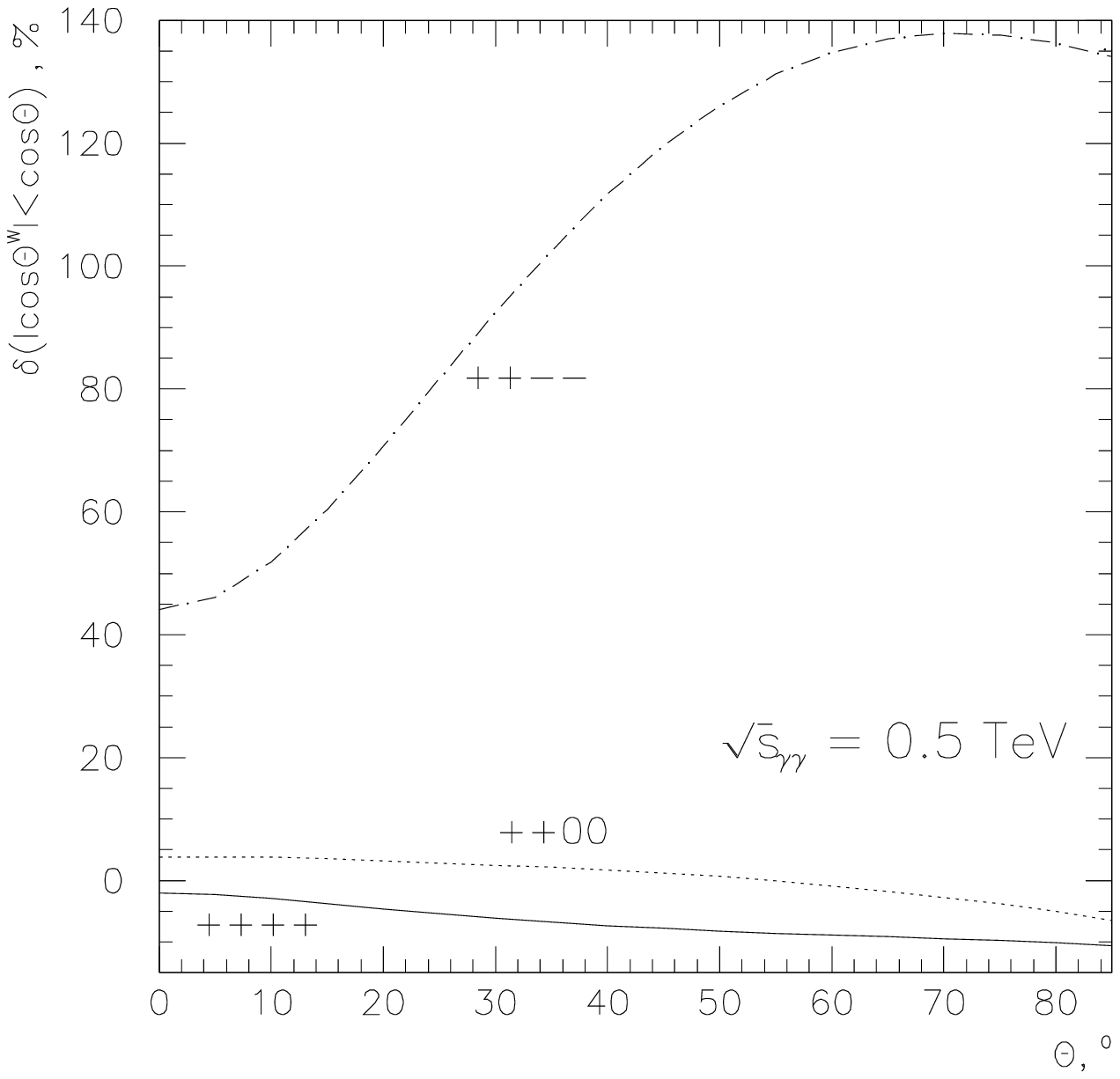,width=3.2in,height=3.5in}}
\put(-.15,0){\epsfig{file=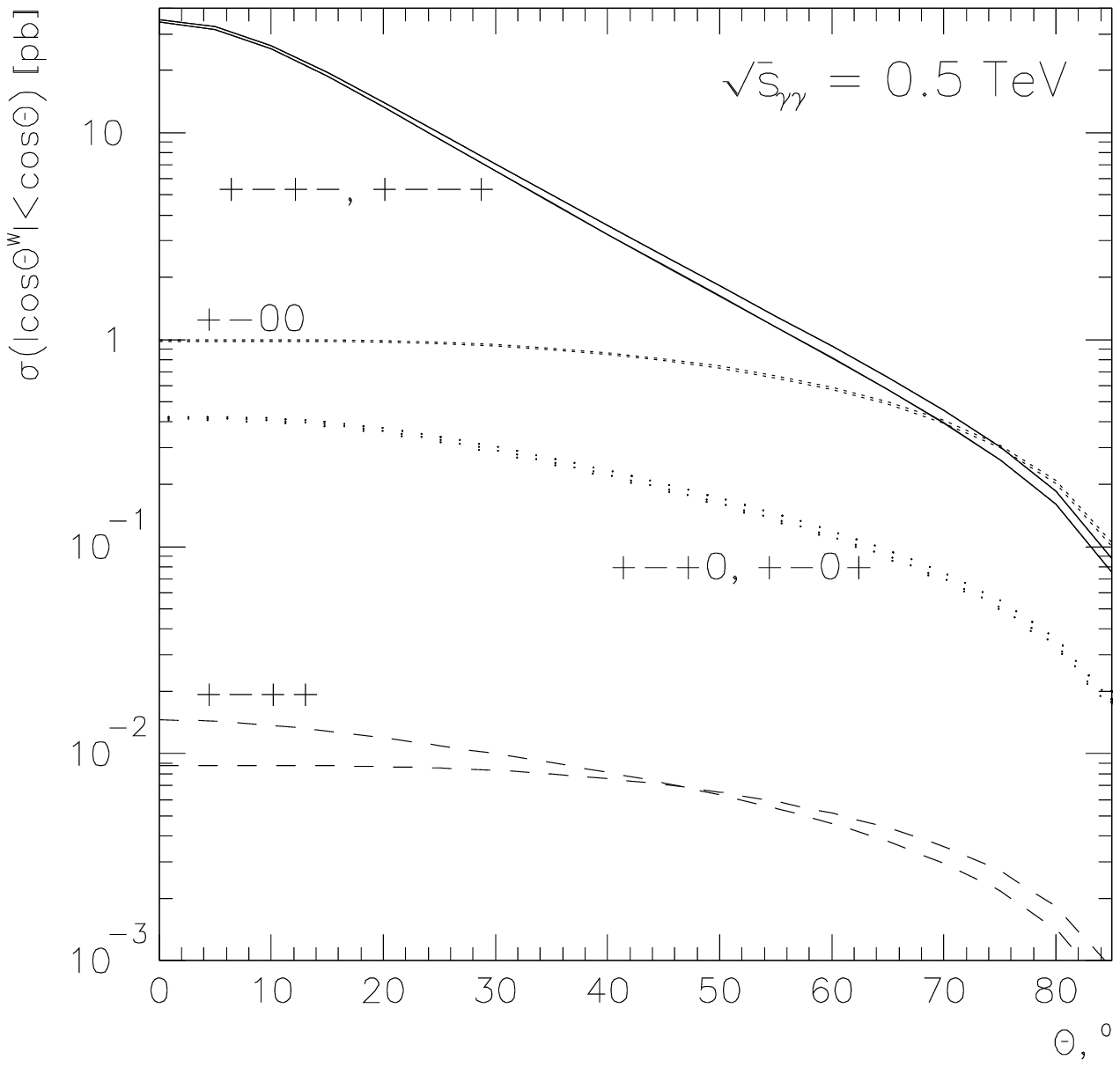,width=3.2in,height=3.5in}}
\put(2.85,0){\epsfig{file=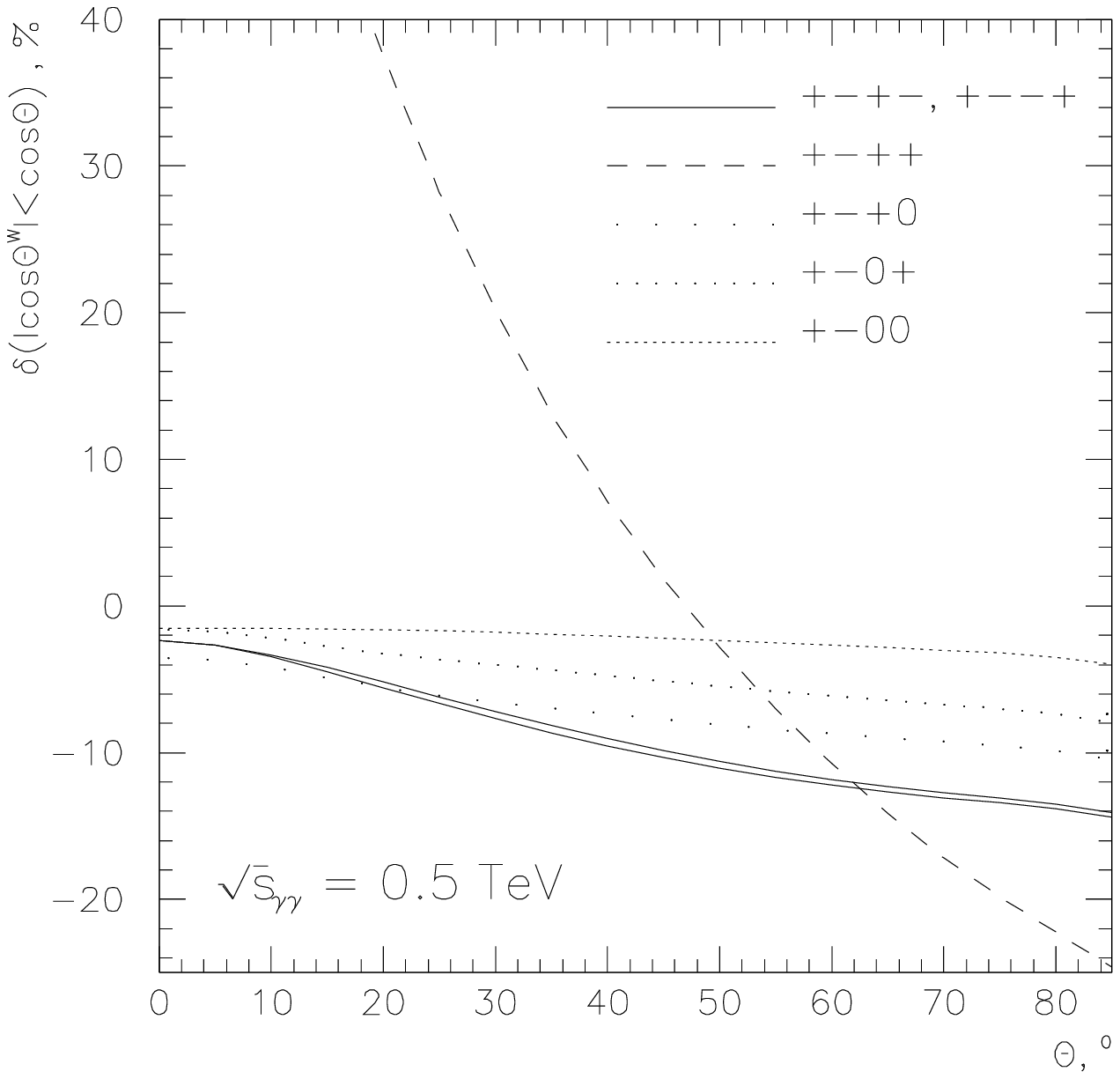,width=3.2in,height=3.5in}}
\end{picture}
\fcaption{Cross sections and relative corrections for $W^+W^-(\g)$
production in the region $|\cos\theta^\pm|<\cos\theta$ for various
polarizations as a function of $\cos\theta$ at
$\sqrt{s_{\g\g}}=0.5$~TeV. Born and corrected cross sections are
shown. The curves nearest to the helicity notations represent the
corrected cross sections.}
\end{figure}

\begin{figure}
\setlength{\unitlength}{1in}
\begin{picture}(6,7)
\put(-.15,3.5){\epsfig{file=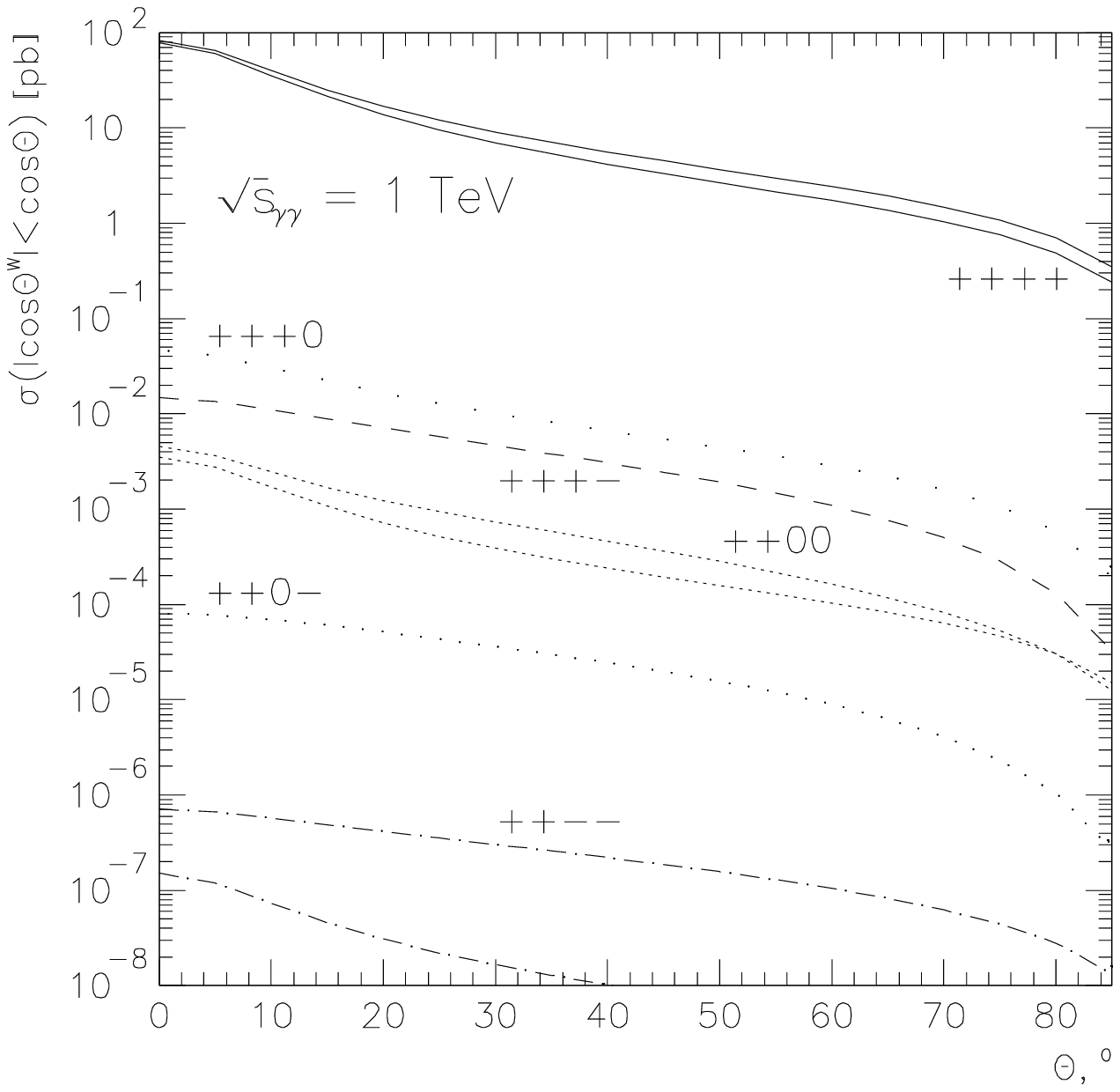,width=3.2in,height=3.5in}}
\put(2.85,3.5){\epsfig{file=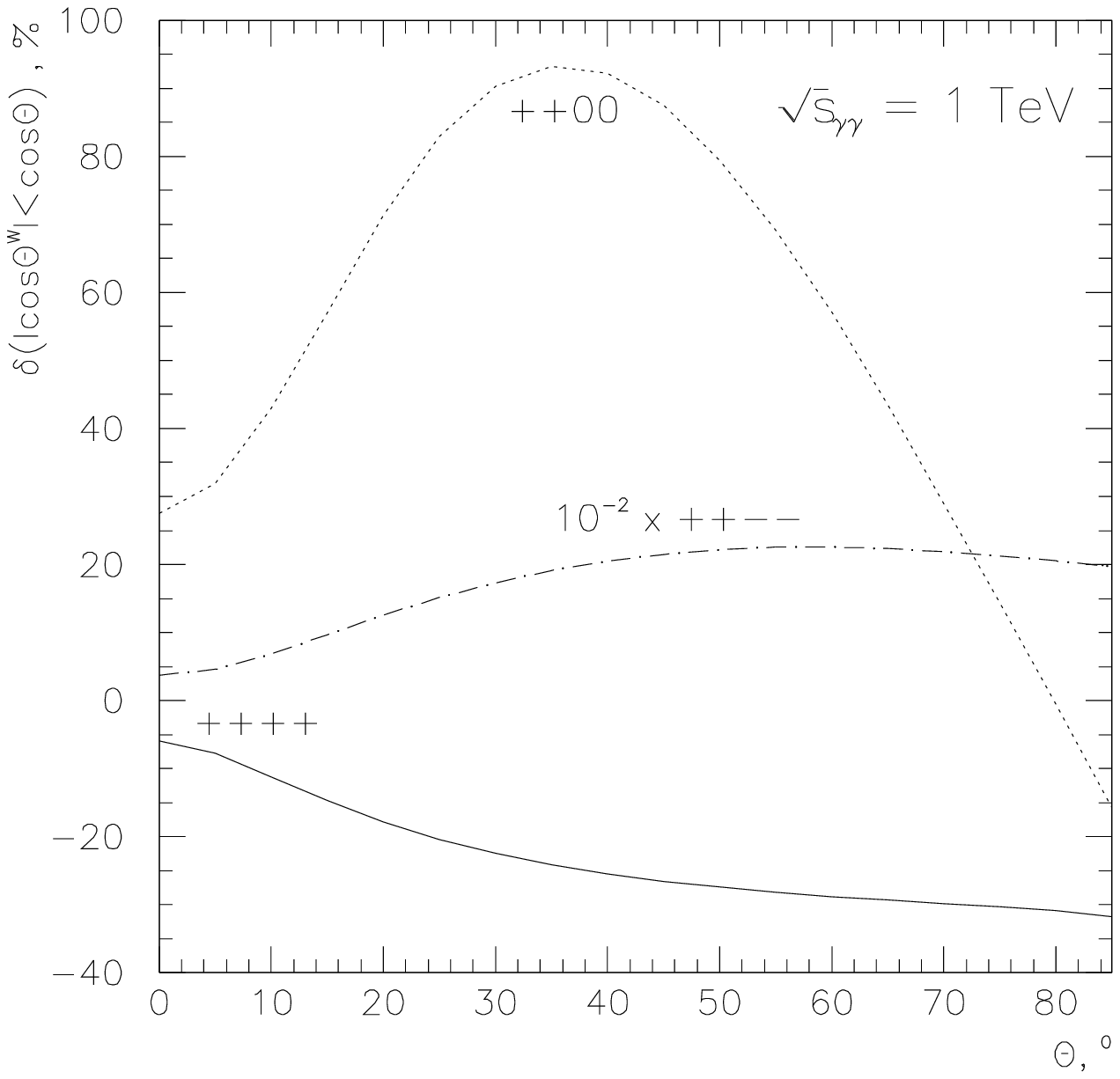,width=3.2in,height=3.5in}}
\put(-.15,0){\epsfig{file=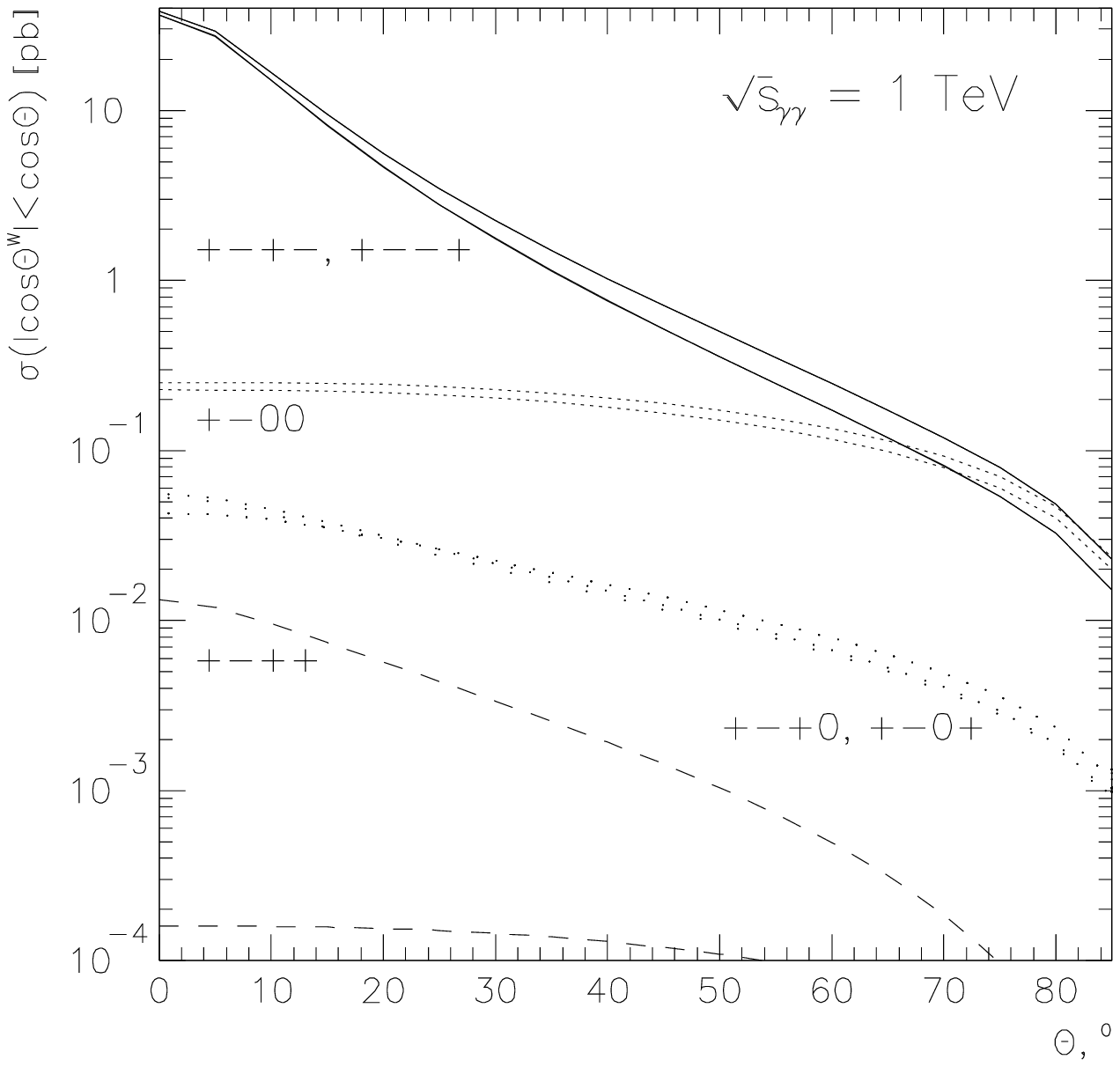,width=3.2in,height=3.5in}}
\put(2.85,0){\epsfig{file=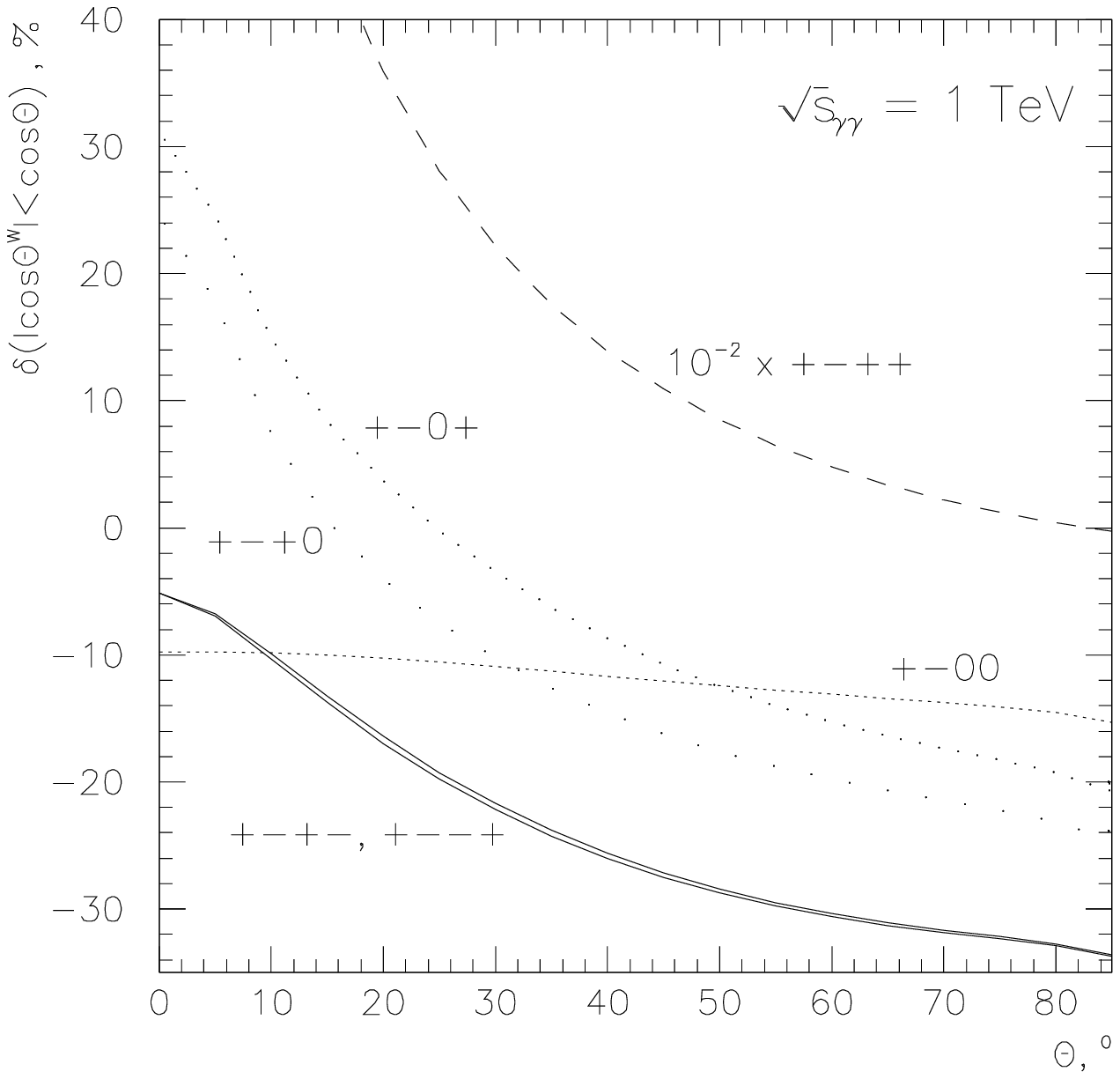,width=3.2in,height=3.5in}}
\end{picture}
\fcaption{The same as Fig.~2 at $\sqrt{s_{\g\g}}=1$~TeV. Since
relative corrections to suppressed at high energy Born cross sections
$\s_{++--},\,\s_{+-++}$ are quite large, corresponding curves
are shown multiplied by a factor of $10^{-2}$.}
\end{figure}

\begin{figure}
\setlength{\unitlength}{1in}
\begin{picture}(6,7)
\put(-.15,3.5){\epsfig{file=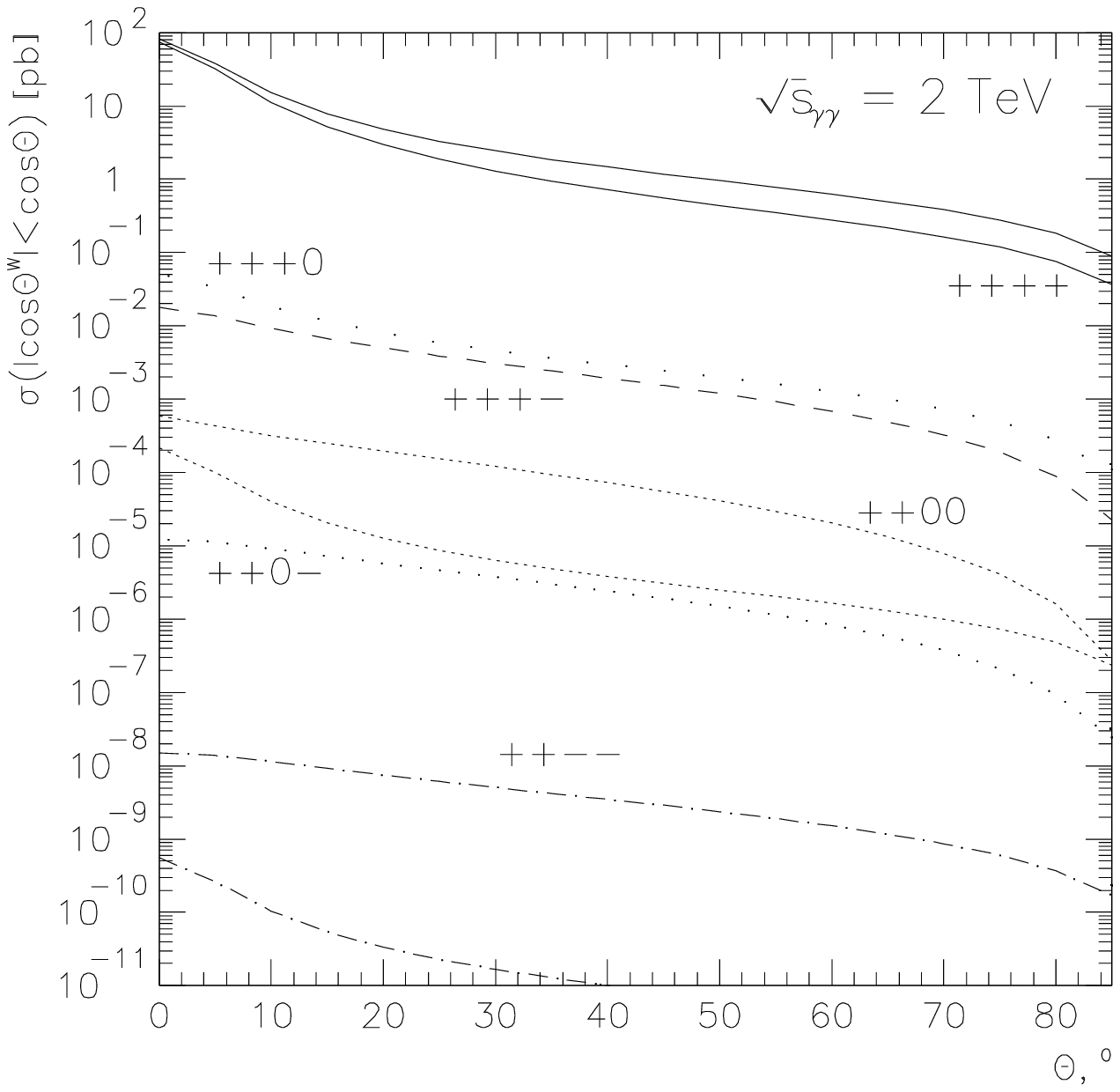,width=3.2in,height=3.5in}}
\put(2.85,3.5){\epsfig{file=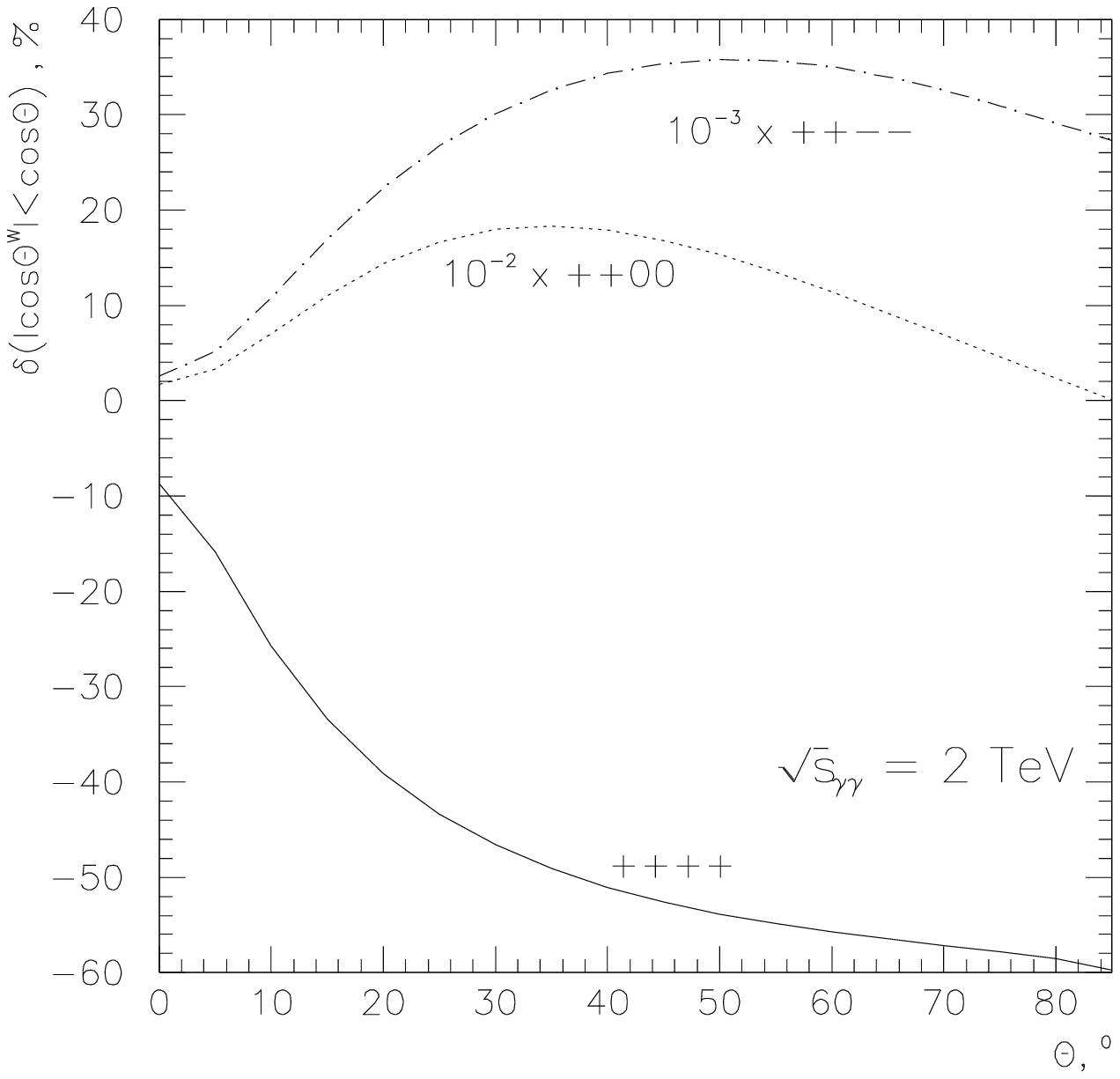,width=3.2in,height=3.5in}}
\put(-.15,0){\epsfig{file=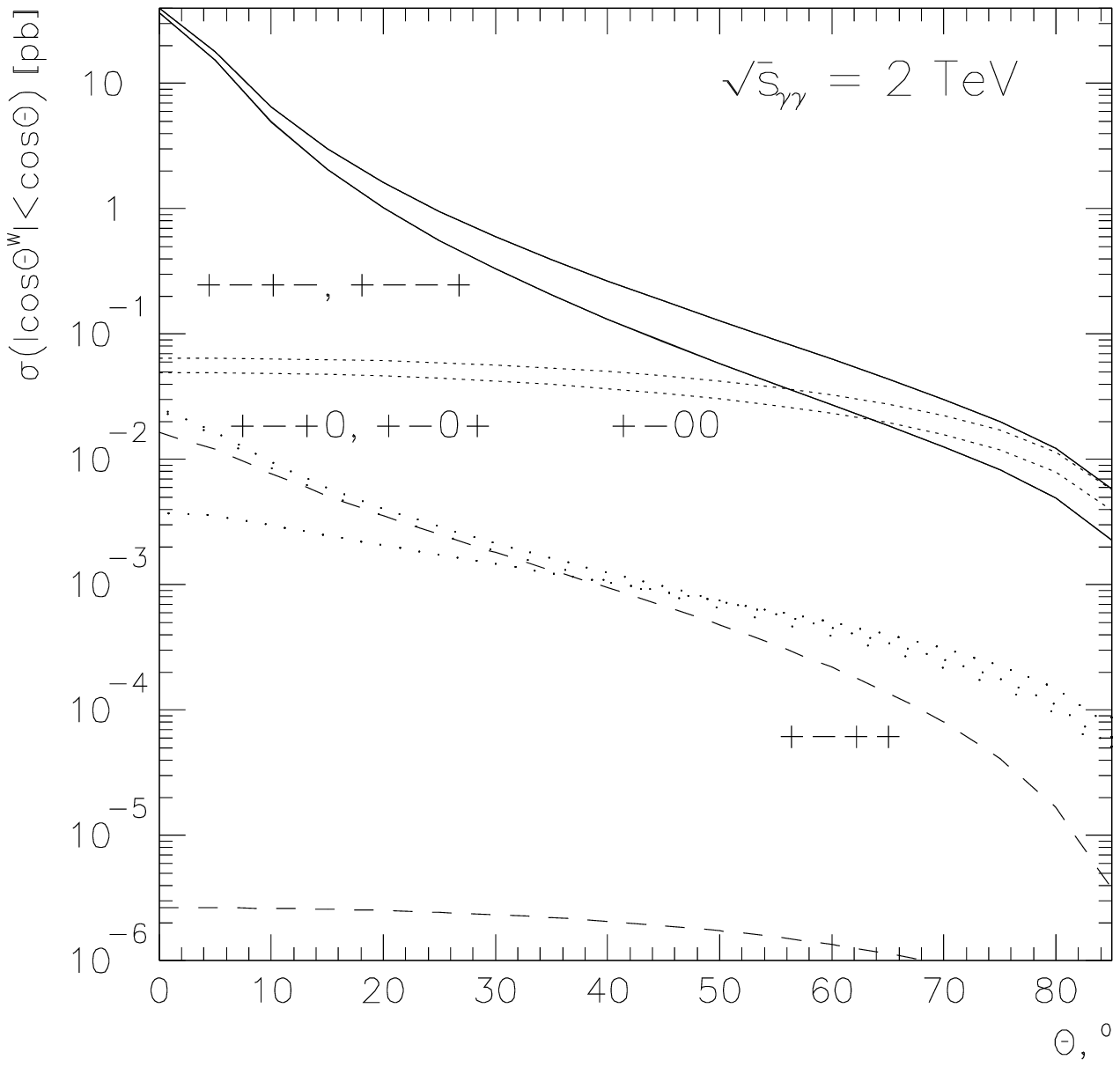,width=3.2in,height=3.5in}}
\put(2.85,0){\epsfig{file=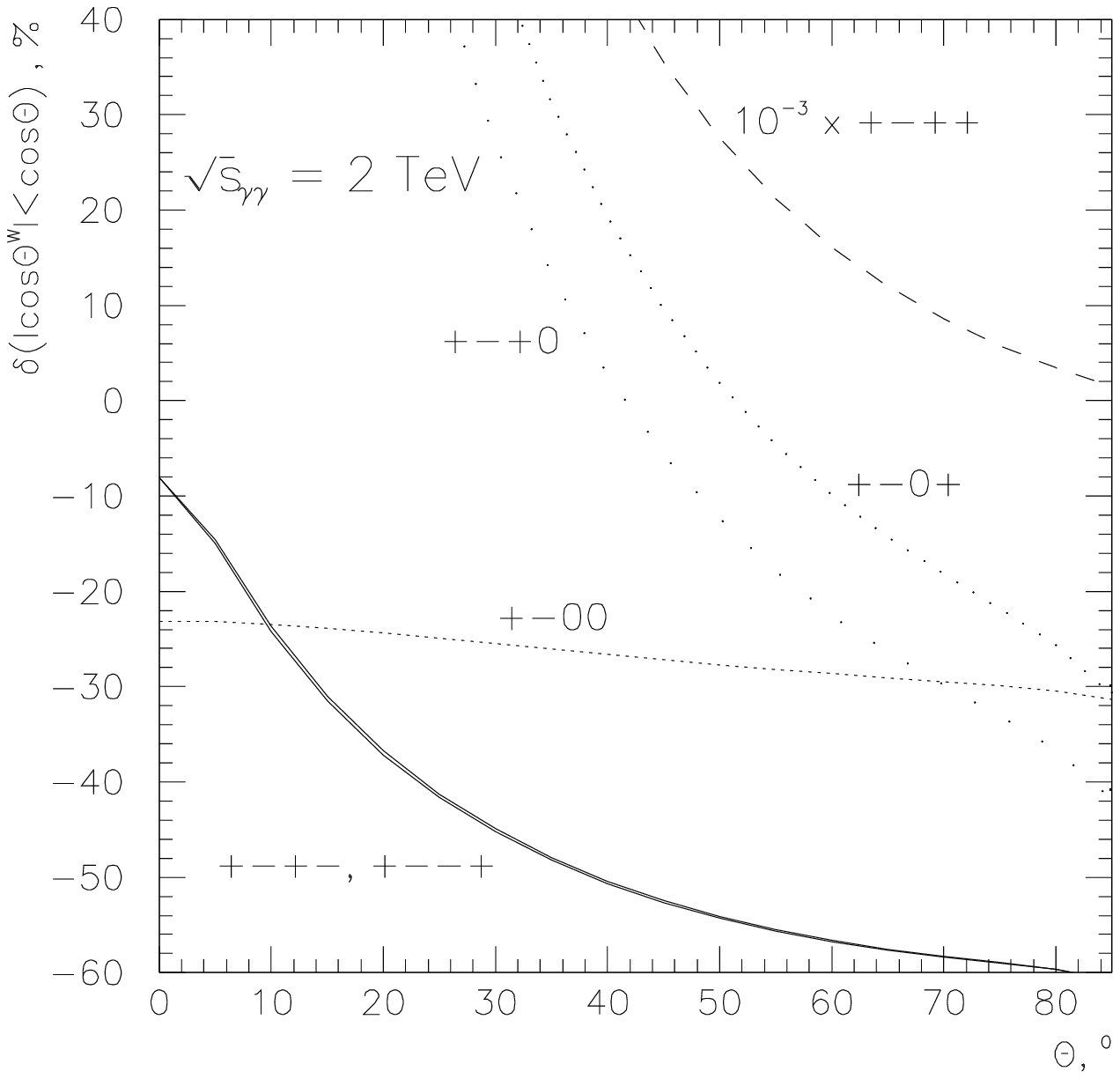,width=3.2in,height=3.5in}}
\end{picture}
\fcaption{The same as Fig.~2 at $\sqrt{s_{\g\g}}=2$~TeV. Relative
corrections are shown rescaled by a factor of $10^{-3}$ for the cross
sections $\s_{++--},\,\s_{+-++}$ and by a factor of $10^{-2}$ for
$\s_{++00}$.}
\end{figure}

Figures~2-4 show the values of the cross sections and ${\cal O}(\a)$
relative corrections for the $WW$-pair production in the central
region  for various helicities at
$\sqrt{s_{\g\g}}=0.5$, 1 and 2~TeV.  As expected, corrections for the
dominating transverse $W_TW_T$-pair production are several times
larger in the central region. As the Born cross section $\s_{++++}$ of
$W^+W^-$ production at $\theta_W=90^\circ$ is asymptotically 16 times
larger than $\s_{+-+-},\,\s_{+--+}$, curves for the cross sections
$\s_{+-+-},\, \s_{+--+}$ fall more rapidly than $\s_{++++}$ and at
$\theta_W=60\div 70^\circ$ the subdominant $\s_{+-00}$ is equal to the
dominant one. The cross sections for all the other helicities are at
least an order of magnitude smaller than the dominant ones in the
whole region of $\cos\theta_W$.

\begin{table}
\tcaption{Total unpolarized Born cross sections and relative
corrections for various intervals of $W^\pm$ scattering
angles. Corrections originating from real hard photon
($\omega_\g>k_c=0.1$~GeV) and $Z$-boson emission as well as IR-finite
sum of soft photon and virtual boson contributions, fermion virtual
corrections and total corrections are given separately.}
\begin{center}
\begin{center}
$\sqrt{s} = 300$~GeV
\end{center}
\begin{tabular}{|c|c|c|c|c|c|c|}\hline
$\theta_{W^\pm}$, ${}^\circ$ & $\sigma^{Born}$, $pb$ & $\delta^{hard}$, \% &
$\delta^{Z}$, \% & $\delta^{soft+bose}$, \% & $\delta^{fermi}$, \% 
&$\delta^{tot}$, \% \\ \hline
$  0^\circ < \theta < 180^\circ$ &  70.22     &  4.15    
& 2.64$\cdot 10^{-2}$& $-$7.09    & 0.327    & $-$1.37    
 \\
$ 10^\circ < \theta < 170^\circ$ &  64.46     &  4.11    
& 2.74$\cdot 10^{-2}$& $-$7.31    & 0.257    & $-$1.59    
 \\
$ 30^\circ < \theta < 150^\circ$ &  38.15     &  4.09    
& 3.27$\cdot 10^{-2}$& $-$8.62    & $-$0.123    & $-$2.67    
 \\
$ 60^\circ < \theta < 120^\circ$ &  12.96     &  4.02    
& 2.94$\cdot 10^{-2}$& $-$10.7    & $-$0.415    & $-$3.75    
\\
\hline   
\end{tabular}
\vspace{.5cm}

\begin{center}
$\sqrt{s} = 500$~GeV
\end{center}
\begin{tabular}{|c|c|c|c|c|c|c|}\hline
$\theta_{W^\pm}$, ${}^\circ$ & $\sigma^{Born}$, $pb$ & $\delta^{hard}$, \% &
$\delta^{Z}$, \% & $\delta^{soft+bose}$, \% & $\delta^{fermi}$, \% 
&$\delta^{tot}$, \% \\ \hline
$  0^\circ < \theta < 180^\circ$ &  77.50     &  7.96    & 0.468    
& $-$10.1    & 9.04$\cdot 10^{-2}$& $-$1.63    
 \\
$ 10^\circ < \theta < 170^\circ$ &  60.71     &  7.89    & 0.541    
& $-$10.7    & $-$0.242    & $-$2.52    
 \\
$ 30^\circ < \theta < 150^\circ$ &  21.85     &  8.05    & 0.817    
& $-$13.0    & $-$1.34    & $-$5.50    
 \\
$ 60^\circ < \theta < 120^\circ$ &  5.681     &  8.02    & 0.789    
& $-$14.8    & $-$2.13    & $-$8.12
\\    
\hline   
\end{tabular}
\vspace{.5cm}

\begin{center}
$\sqrt{s} = 1000$~GeV
\end{center}
\begin{tabular}{|c|c|c|c|c|c|c|}\hline
$\theta_{W^\pm}$, ${}^\circ$ & $\sigma^{Born}$, $pb$ & $\delta^{hard}$, \% &
$\delta^{Z}$, \% & $\delta^{soft+bose}$, \% & $\delta^{fermi}$, \% 
&$\delta^{tot}$, \% \\ \hline
$  0^\circ < \theta < 180^\circ$ &  79.99     &  13.3    &  1.55    
& $-$18.7    & $-$5.51$\cdot 10^{-2}$& $-$3.89    
 \\
$ 10^\circ < \theta < 170^\circ$ &  37.04     &  13.4    &  2.39    
& $-$22.6    & $-$1.28    & $-$8.10    
 \\
$ 30^\circ < \theta < 150^\circ$ &  6.924     &  14.2    &  3.96    
& $-$32.1    & $-$3.80    & $-$17.8    
 \\
$ 60^\circ < \theta < 120^\circ$ &  1.542     &  14.2    &  3.88    
& $-$37.1    & $-$5.13    & $-$24.1    
 \\
\hline   
\end{tabular}
\vspace{.5cm}

\begin{center}
$\sqrt{s} = 2000$~GeV
\end{center}
\begin{tabular}{|c|c|c|c|c|c|c|}\hline
$\theta_{W^\pm}$, ${}^\circ$ & $\sigma^{Born}$, $pb$ & $\delta^{hard}$, \% &
$\delta^{Z}$, \% & $\delta^{soft+bose}$, \% & $\delta^{fermi}$, \% 
&$\delta^{tot}$, \% \\ \hline
$  0^\circ < \theta < 180^\circ$ &  80.53     &  19.0    &  2.91    
& $-$27.2    & $-$7.45$\cdot 10^{-2}$& $-$5.33    
 \\
$ 10^\circ < \theta < 170^\circ$ &  14.14     &  20.1    &  6.38    
& $-$41.6    & $-$2.99    & $-$18.1    
 \\
$ 30^\circ < \theta < 150^\circ$ &  1.848     &  21.5    &  9.77    
& $-$60.1    & $-$6.54    & $-$35.4    
 \\
$ 60^\circ < \theta < 120^\circ$ & 0.3936     &  21.6    &  9.60    
& $-$67.6    & $-$8.04    & $-$44.5    
\\
\hline   
\end{tabular}
\end{center}
\end{table}

In Table~2 Born cross sections and relative corrections are given for
several intervals of $W^\pm$ scattering angles. At high energies large
cancellations occur between negative virtual corrections and positive
corrections corresponding to real photon or $Z$-boson emission, as one
should expect, as at high energies, $s\gg M_W^2$, collinear
singularities should cancel out only in the sum of virtual, soft and
hard contributions again as a consequence of KLN
theorem. Consequently, although the correction originating from the
$WWZ$ production is completely negligible at
$\sqrt{s_{\g\g}}=0.3$~TeV, it is of the same order of magnitude as
hard photon correction at 2~TeV. The size of the total corrections is
in rough agreement with the values expected on the basis of leading
logarithmic corrections discussed at the end of the
Section~5. Corrections in the central region $60^\circ < \theta <
120^\circ$ are really $5\div 8$ times larger than the corrections to
the total cross section at $0.5\div 2$~TeV.

\begin{figure}
\setlength{\unitlength}{1in}
\begin{picture}(6,3.5)
\put(1.35,0){\epsfig{file=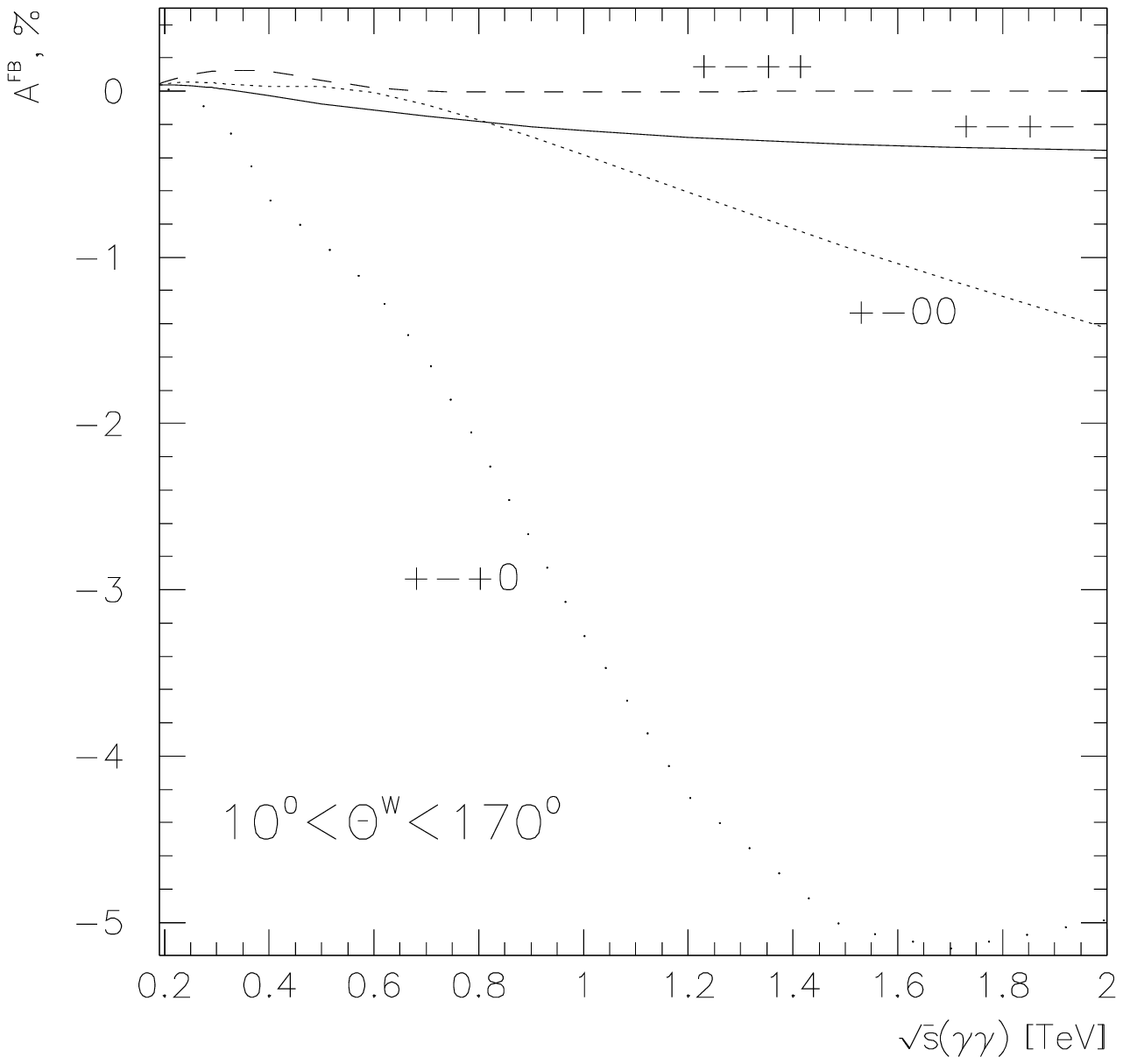,width=3.2in,height=3.5in}}
\end{picture}
\fcaption{$P$-odd forward-backward asymmetry as a function of the
center-of-mass energy.}
\end{figure}

Although the fermionic one-loop corrections are small, only they give
nonvanishing contribution to $P$- or $C$-violating asymmetries. As an
example of such a quantity we consider the $P$-odd forward-backward
asymmetry 
\beq 
A_{\l_3\l_4}^{FB} =
\frac{\s_{+-\l_3\l_4}(\cos\theta^{W^+}>0) -
\s_{+-(-\l_3)(-\l_4)}(\cos\theta^{W^+}<0)
}{\s_{+-\l_3\l_4}(\cos\theta^{W^+}>0) +
\s_{+-(-\l_3)(-\l_4)}(\cos\theta^{W^+}<0)} 
\eeq
Figure~5 presents the forward-backward asymmetry for various
helicities as a function of c.m.s. energy. For the dominant transverse
$\s_{+-+-},\,\s_{+--+}$ cross sections the asymmetry at high energy is
negative and less than 0.5\%. Even for the subdominant $\s_{+-00}$
cross section the asymmetry is negative and smaller that 1.5\% up to
2~TeV. For the suppressed cross sections $\s_{+-+0},\, \s_{+--0}$ the
fermionic contribution gives rise to the asymmetry of about $-5\%$ at
high energy.

Another example of the asymmetry, that could be used for the
measurement of the $W$-boson anomalous magnetic and quadrupole
moments\cite{BrodskyRizzoSchmidt} is given by the so called
polarization asymmetry
\beq
A^{+-}=
\frac
{\sum_{\l_i}(\s_{++\l_3\l_4(\l_5)} - \s_{+-\l_3\l_4(\l_5)})}
{\sum_{\l_i}(\s_{++\l_3\l_4(\l_5)} + \s_{+-\l_3\l_4(\l_5)})}.
\label{asym_pm}
\eeq

\begin{figure}
\setlength{\unitlength}{1in}
\begin{picture}(6,3.5)
\put(1.35,0){\epsfig{file=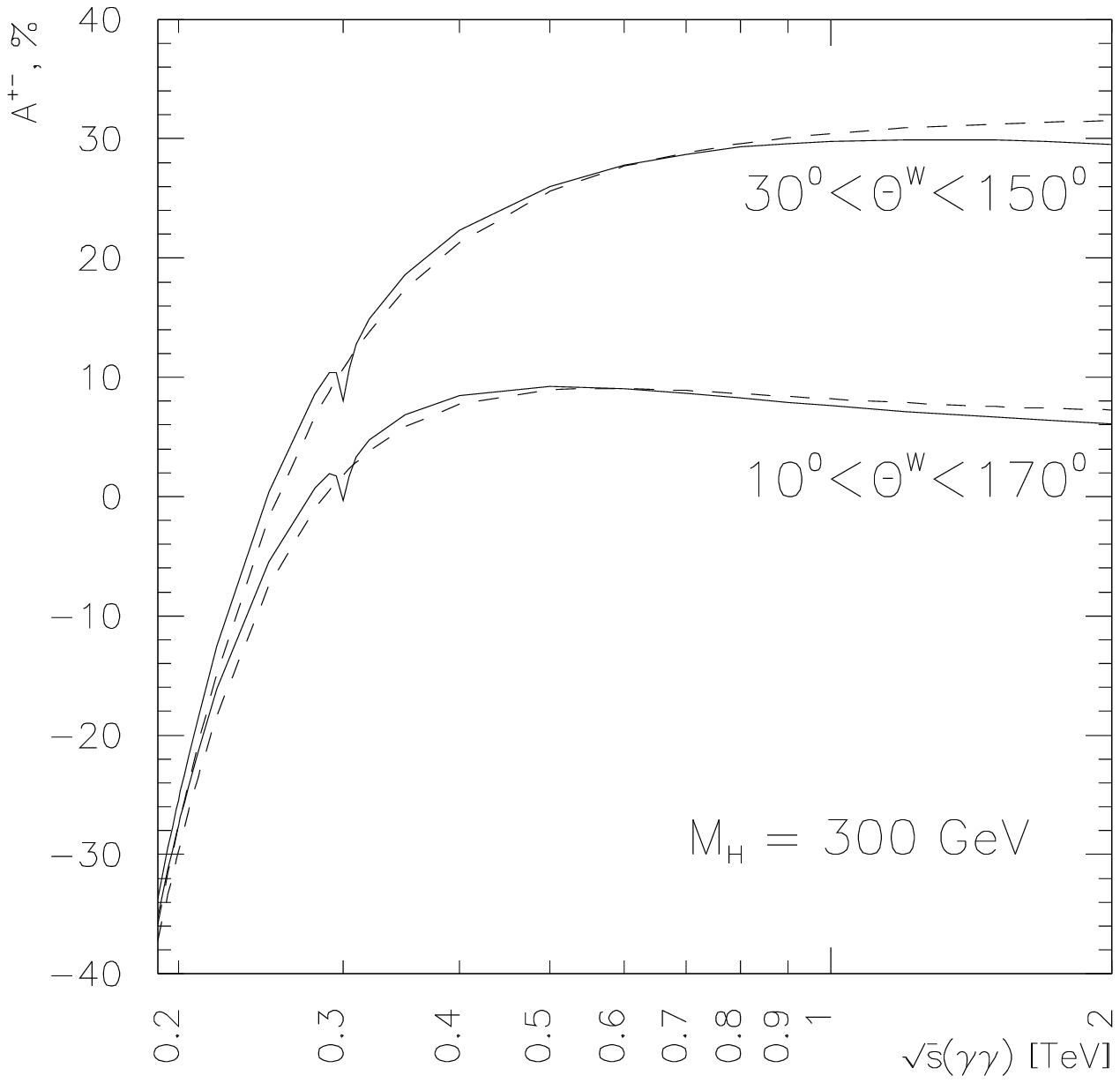,width=3.2in,height=3.5in}}
\end{picture}
\fcaption{Polarization asymmetry as a function of the center-of-mass
energy. Dashed line is Born asymmetry and solid curve is the asymmetry
with account of the radiative corrections.}
\end{figure}

Using a quantum loop expansion it was shown that the logarithmic
integral of the spin-dependent photoabsorption cross section
$\int^\infty_{\omega_{th}} {d\omega_\g\over \omega_\g} \Delta
\sigma^{Born}(\omega_\g)$ vanishes for any $2 \to 2$ SM
process $\gamma a \to b c$ in the classical, tree-graph approximation
\cite{BrodskyRizzoSchmidt}. Here $\Delta\s = \s_{++}- \s_{+-}$ is the
difference between the photoabsorption cross section for parallel and
antiparallel photon and target helicities. A mean value theorem then
implies that there must be a center of mass energy where the
polarization asymmetry possesses a zero. The position of the zero may
be determined with sufficient precision to constrain the anomalous
couplings of the $W$ to better than the 1\% level at $95\%$ CL
\cite{BrodskyRizzoSchmidt}.  Figure~6 shows the polarization asymmetry
(\ref{asym_pm}) for two different angular cuts as a function of
energy.  Radiative corrections shift the position of zero at about
280~GeV for the angular cut $10^\circ < \theta^\pm < 170^\circ$
(250~GeV for $30^\circ < \theta^\pm< 150^\circ$) to a lower energy
by about 10~GeV. The dip from the 300~GeV Higgs boson is clearly
seen. The corrections to the polarization asymmetry are of the order
of $5\div 10\%$ (outside the region near the crossing point) and so
should necessarily be taken into account in realistic determinations
of the precise constraints on the anomalous couplings of the $W$.

\begin{figure}
\setlength{\unitlength}{1in}
\begin{picture}(6,7)
\put(-.15,3.5){\epsfig{file=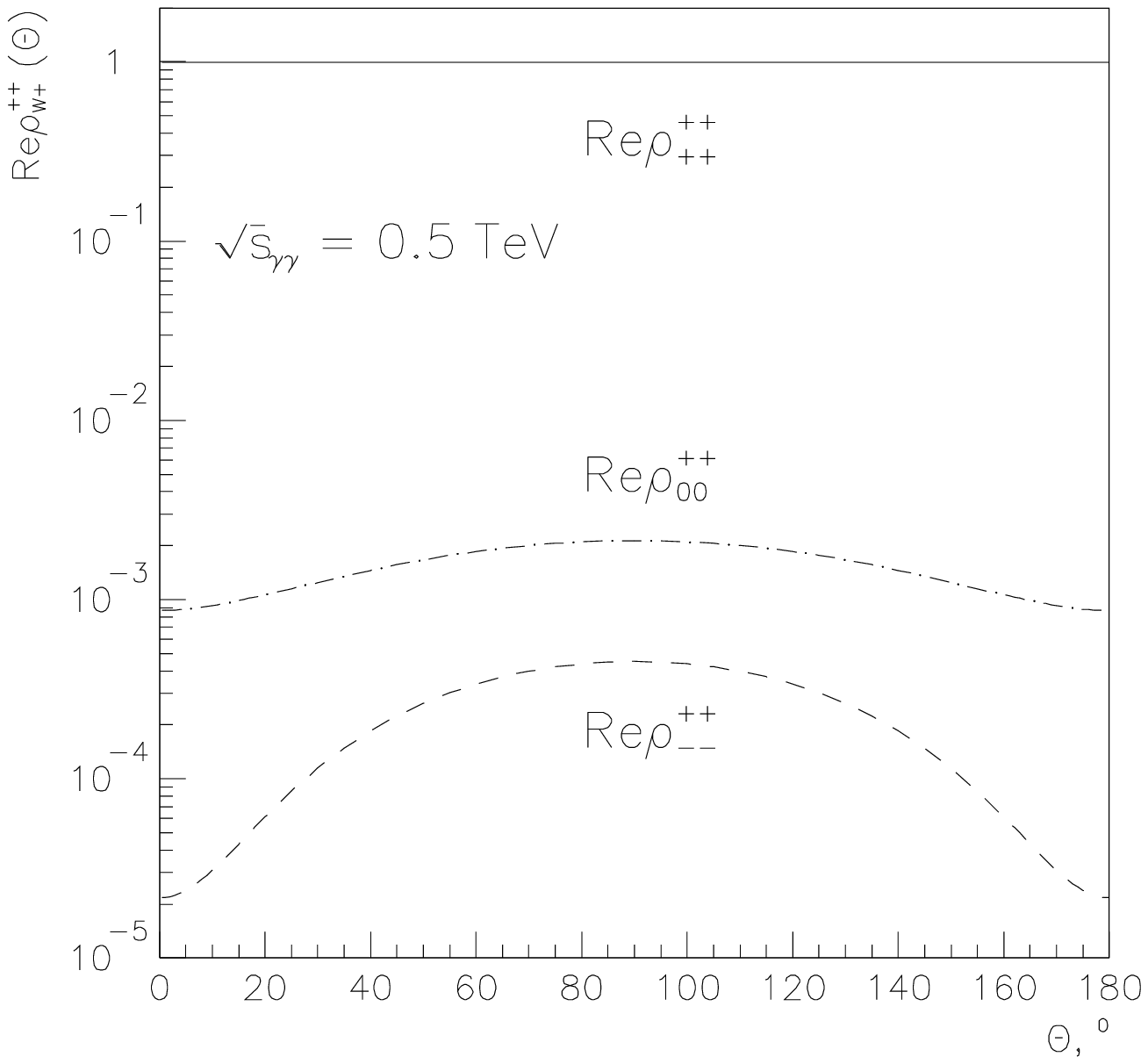,width=3.2in,height=3.5in}}
\put(2.85,3.5){\epsfig{file=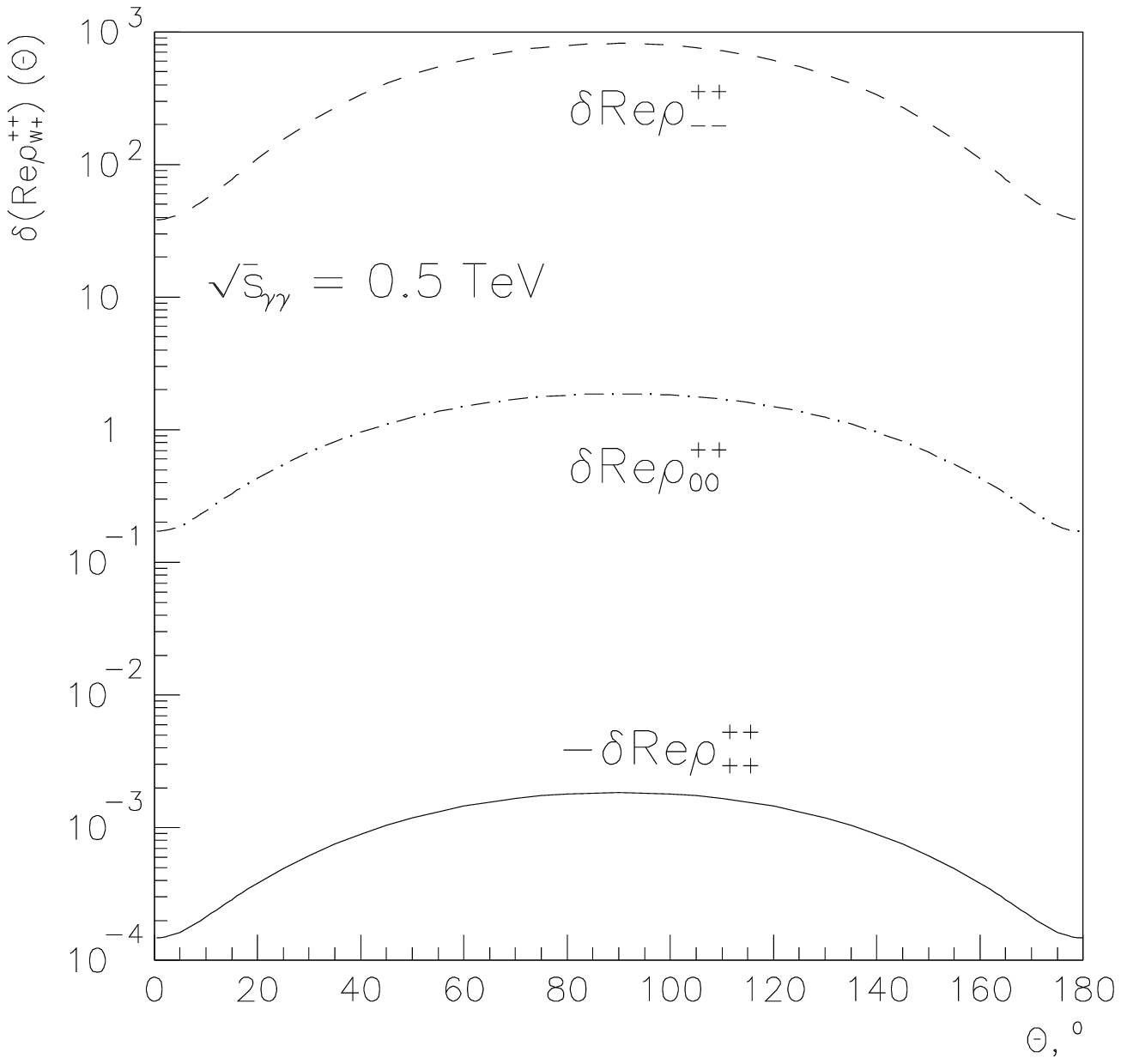,width=3.2in,height=3.5in}}
\put(1.35,0){\epsfig{file=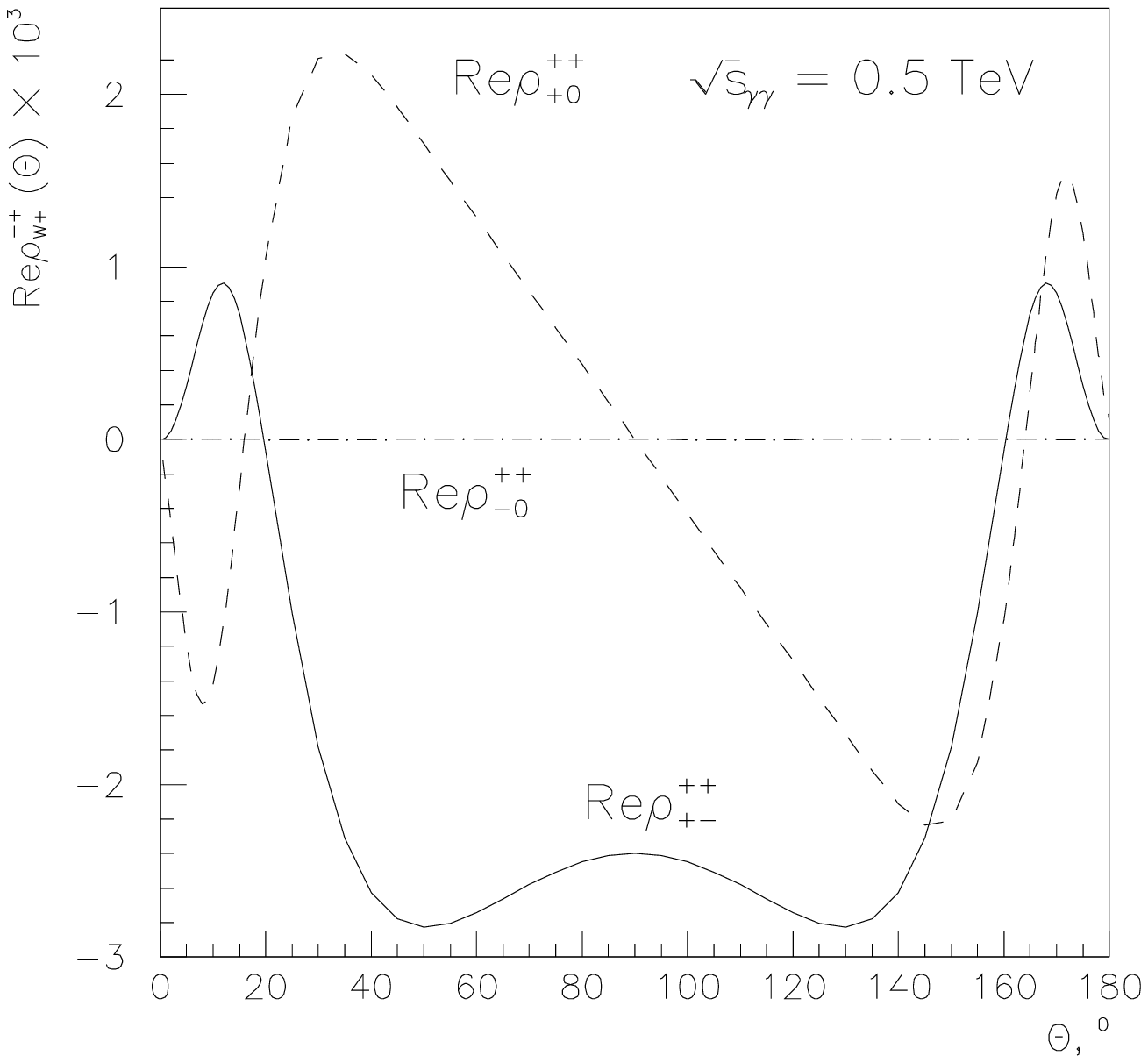,width=3.2in,height=3.5in}}
\end{picture}
\fcaption{The dependence on the production angle of the corrected
single-particle density matrix elements and relative corrections for
equal initial photons helicities $\l_1=\l_2=1$.}
\end{figure}

\begin{figure}
\setlength{\unitlength}{1in}
\begin{picture}(6,7)
\put(-.15,3.5){\epsfig{file=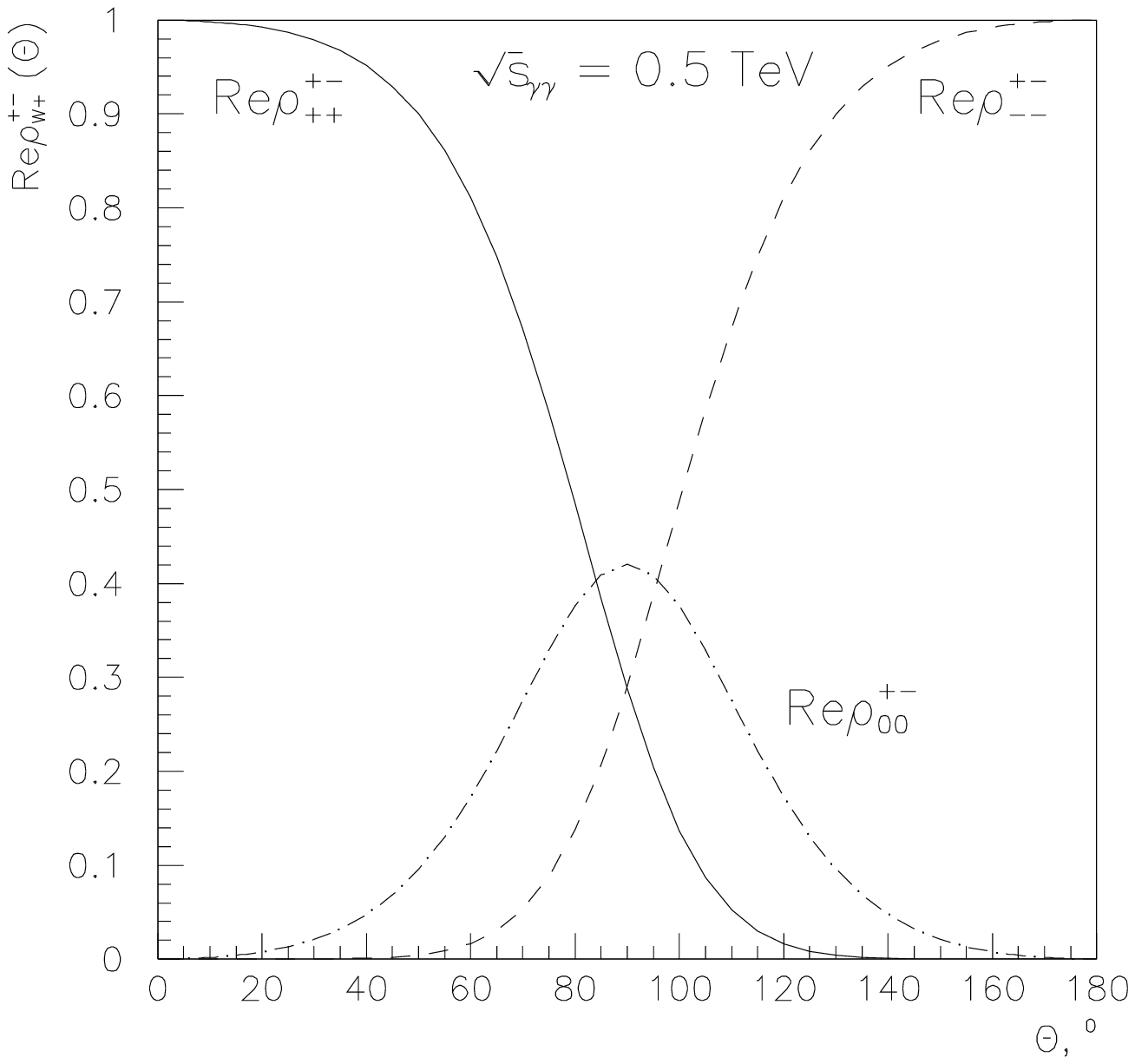,width=3.2in,height=3.5in}}
\put(2.85,3.5){\epsfig{file=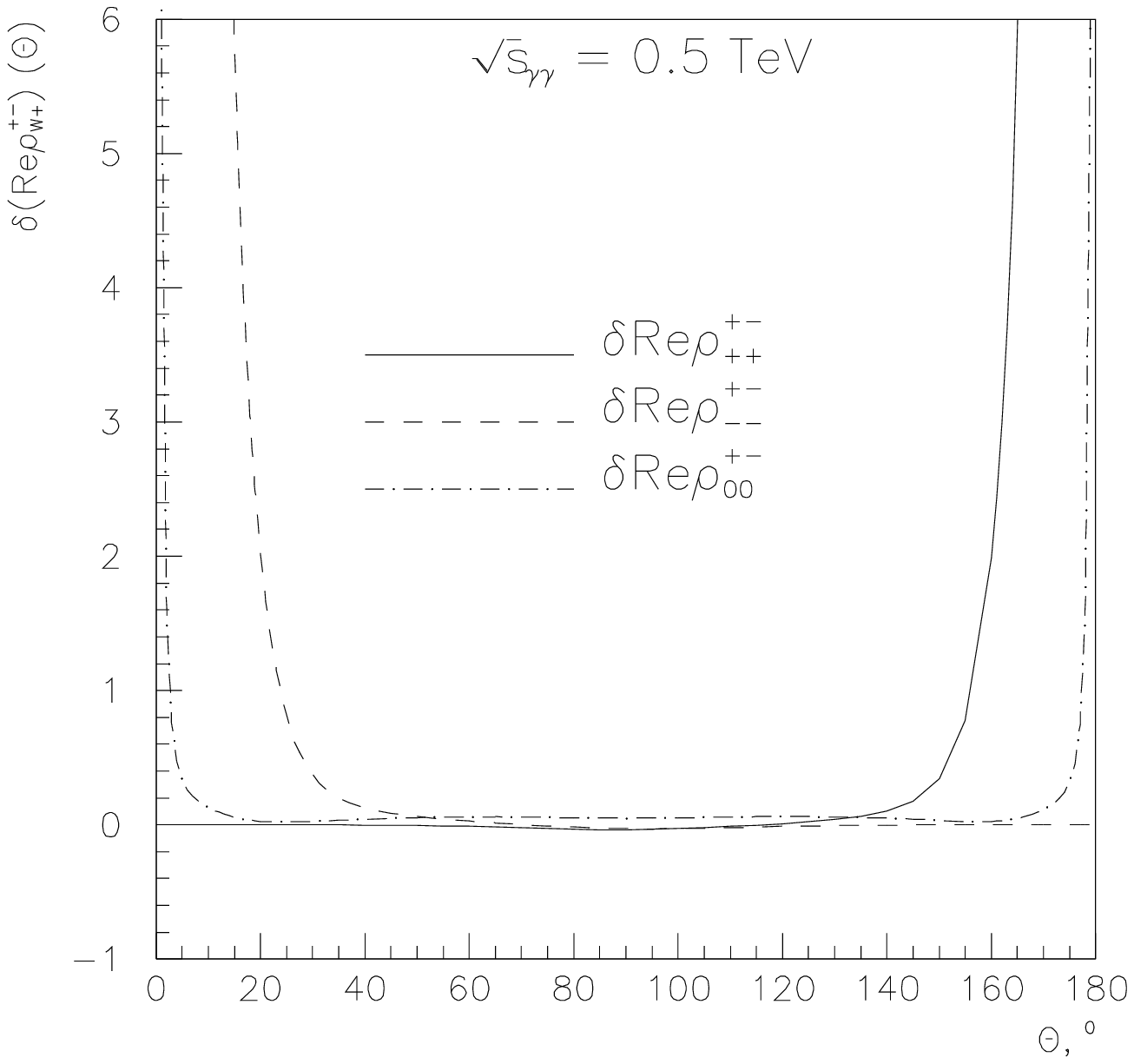,width=3.2in,height=3.5in}}
\put(-.15,0){\epsfig{file=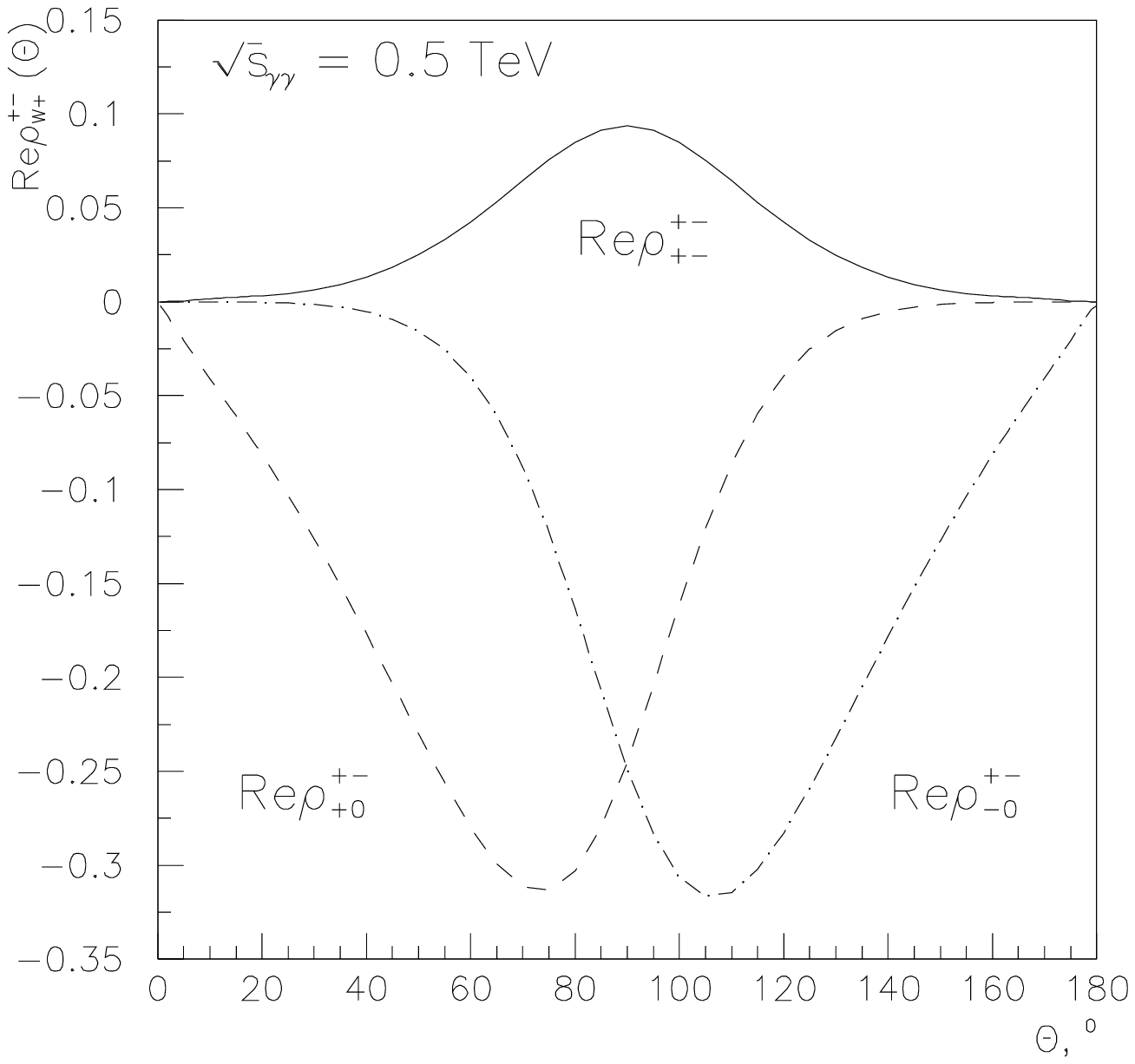,width=3.2in,height=3.5in}}
\put(2.85,0){\epsfig{file=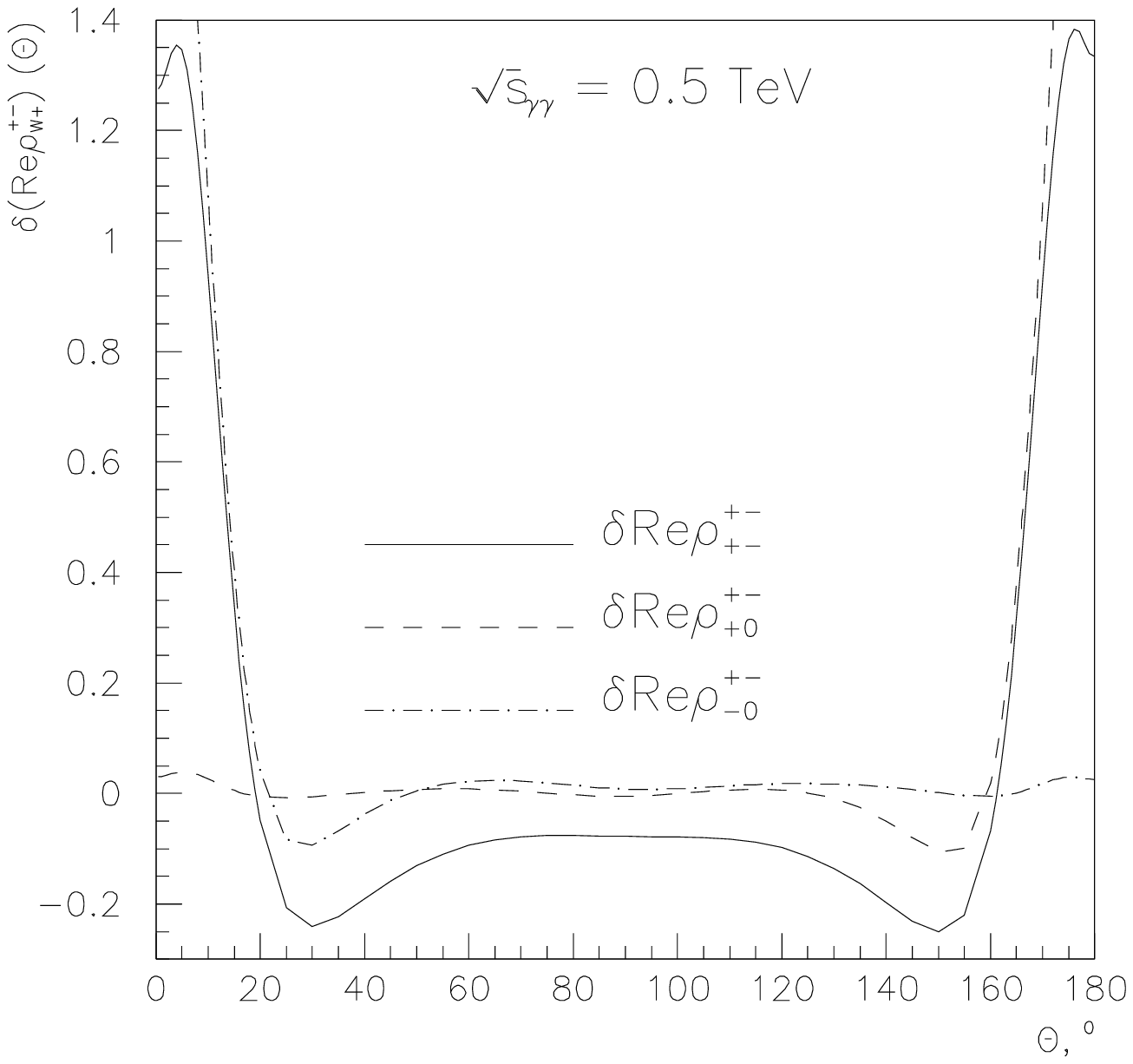,width=3.2in,height=3.5in}}
\end{picture}
\fcaption{The dependence on the production angle of the corrected
single-particle density matrix elements and relative corrections for
opposite initial photons helicities $\l_1=-\l_2=1$.}
\end{figure}

\begin{figure}
\setlength{\unitlength}{1in}
\begin{picture}(6,3.5)
\put(-.15,0){\epsfig{file=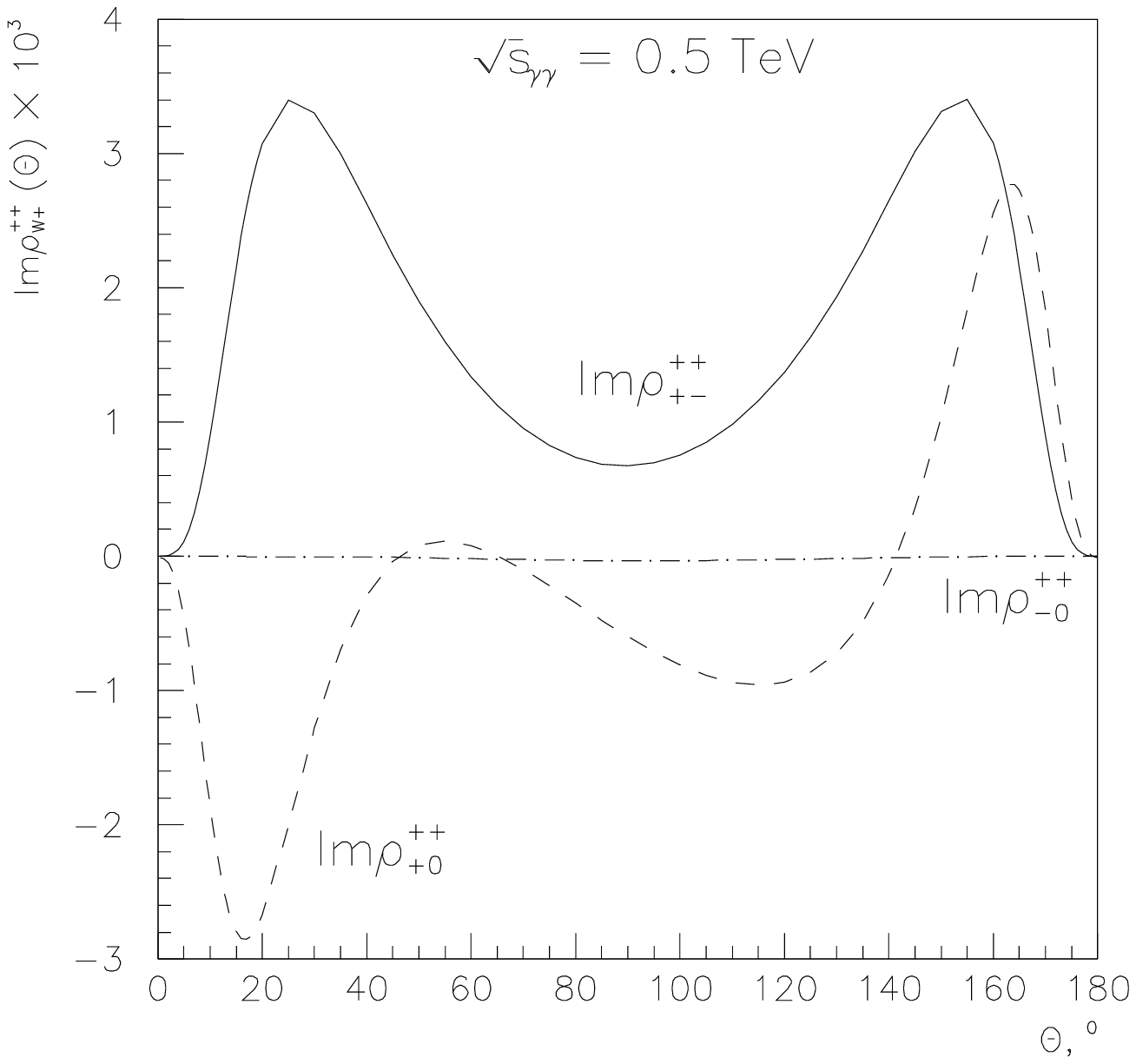,width=3.2in,height=3.5in}}
\put(2.85,0){\epsfig{file=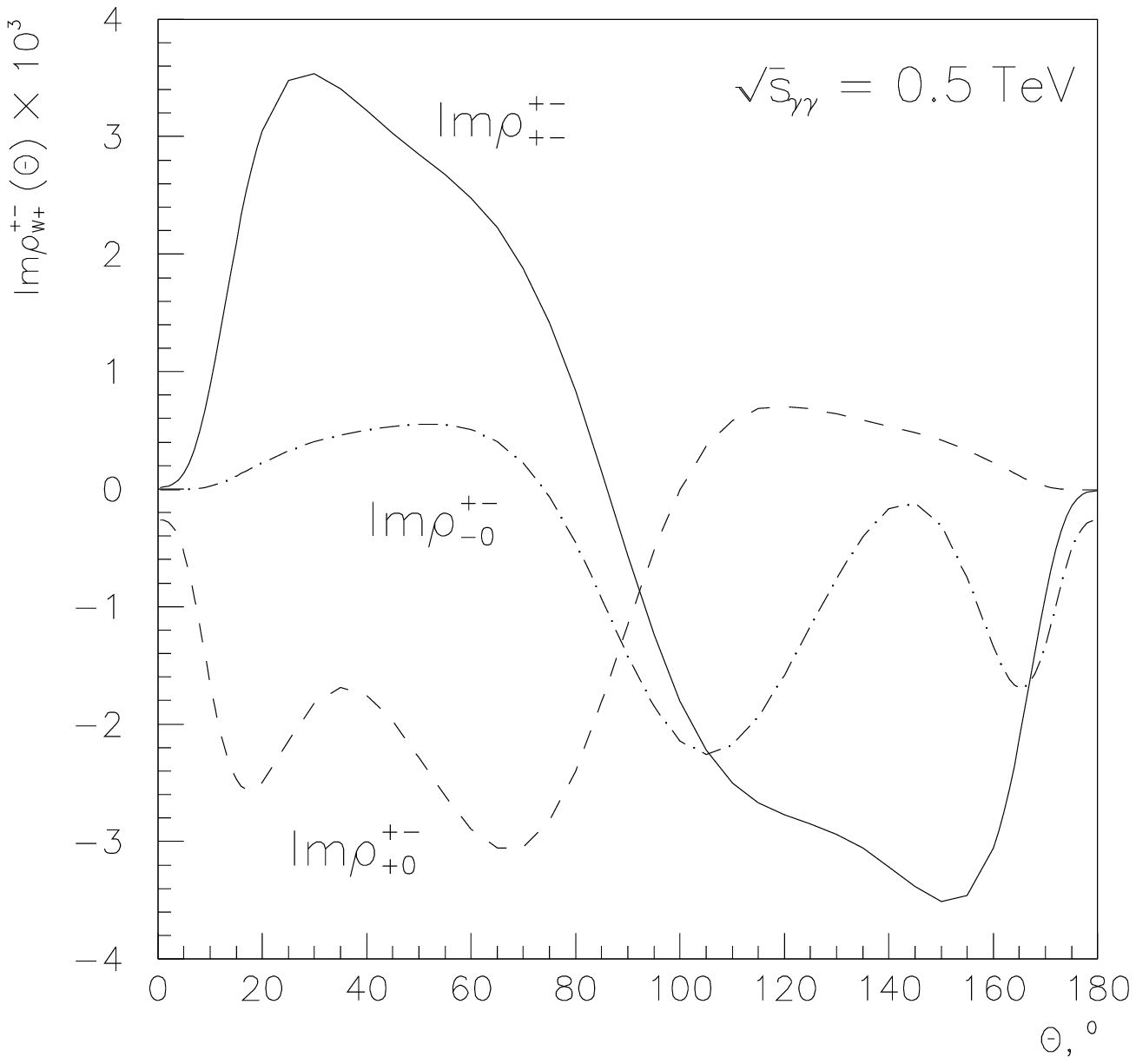,width=3.2in,height=3.5in}}
\end{picture}
\fcaption{The dependence on the production angle of the
imaginary parts of the single-particle density matrix elements.}
\end{figure}

The most efficient way to experimentally constrain the anomalous
couplings of the $W$-boson requires the extraction of the $W^\pm$
spin-density matrix and the $W^+$, $W^-$ spin correlations from
future data \cite{gamma-gamma2,BBB,BM2}. The polarization properties of
the single $W^+$ (or $W^-$) decays are described by the
single-particle density matrices
\bea
&&{\rho^{\l_1\l_2}_{\l_3\l_3'}}_{W^+}(\theta^+)
 \propto 
\sum_{\l_4}{\cal M}_{\l_1\l_2\l_3\l_4}(\g\g\to W^+W^-)
{\cal M}^*_{\l_1\l_2\l_3'\l_4}(\g\g\to W^+W^-)dPS^{(2)}
\nonumber\\
&& + \sum_{\l_4,\l_5}\int 
{\cal M}_{\l_1\l_2\l_3\l_4\l_5}(\g\g\to W^+W^-\g)
{\cal M}^*_{\l_1\l_2\l_3'\l_4\l_5}(\g\g\to W^+W^-\g)dPS^{(3)},
\label{density}
\eea
here $\theta^+$ is the $W^+$ production angle in the c.m.s.
The normalization is
\beq
\sum_{\l_3,\l_3'}{\rho^{\l_1\l_2}_{\l_3\l_3'}}_{W^+}(\theta^+)=1.
\eeq

In Figures~7--9 we show the $W^+$ density-matrix elements
(\ref{density}) at the $\g\g$ center-of-mass energy of 500~GeV. Also
the relative corrections with respect to the tree-level values are
shown.  Since off-diagonal matrix elements for equal initial photon
helicities vanish at the tree-level, no relative corrections are
given for these in the Figure~7. The imaginary parts shown in the
Figure~9 also first appear at the one-loop level. Corrections to the
dominant matrix elements are always less than 1\%. The corrections to
the subdominant matrix elements or to the dominant ones at the angles
where they are suppressed are be quite large.

\section{Conclusions}

The high energy laser induced $\g\g$ collider will be a real $W$
factory and an ideal laboratory for precision tests on anomalous $W$
interactions.  The theoretical predictions for $W$ pair production,
including complete electroweak ${\cal O}(\a)$ SM radiative
corrections, are obtained with very little theoretical uncertainty at
least for energies below 1~TeV. It is absolutely
necessary to take them into account for precise measurements of the
$W$ properties.  

Of course, searches for a new interactions
will be the most interesting part of future experiments with $W$ pair
production. However, it is worth mentioning that $WW$ production in
photon-photon collisions should also qualify as a good $\g\g$
luminosity monitor \cite{BB,BBB,luminosity}. As the interaction region in
photon-photon collisions is quite complicated \cite{gg,balakin,schulte}, it
will be extremely important to measure the $\g\g$ differential luminosity
and reconstruct the polarized photon spectra.
In fact, one can imagine that the reaction $\g\g\to W^+W^-$ could be
used both to uncover new physics and to calibrate luminosity. Indeed,
anomalies would affect longitudinal and central $W$'s, while to
measure luminosity one would use most copiously produced transverse
and forward/backward $W$'s. Moreover, the analysis of Section~5 of the
leading heavy Higgs boson and top-quark contributions shows, that the
dominating cross sections of forward/backward transverse $W_TW_T$ pair
production are essentially insensitive to the details of new
physics. Table~2 shows that radiative corrections for forward $W$
scattering are quite small even at 2~TeV.

From the other side, as corrections for the total cross section in the
central region are as large as $-(20\div 40)\%$ at 2 TeV, a very
relevant question relates to the size of the higher loop
corrections. As the structure of leading corrections at high energies
is known to have Regge form, one can imagine that at very high
energies it would be possible to experimentally study the reggeization
properties of massive weak bosons which are a unique feature of
non-abelian spontaneously broken gauge theory.

\section*{Acknowledgements}

This work was supported in part by the Alexander von Humboldt 
Foundation and the Russian Foundation for Basic Research grant 96-02-19-464.

\end{document}